\documentclass[printer]{aa}

\usepackage{graphicx}
\usepackage{txfonts}
\begin{document} 

   \title{Measuring the physical imprints of gas flows in galaxies I: Accretion rate histories}
   \titlerunning{Accretion rate histories}

   \author{A. Camps-Fariña
          \inst{1,2}\fnmsep\thanks{\email{arcamps@ucm.es}},
          P. S\'{a}nchez-Blázquez\inst{1,2},
          S. Roca-F\`{a}brega\inst{1,2,3}
          \and
          S. F. S\'{a}nchez\inst{4}
          }
    \authorrunning{A. Camps-Fariña et al.}

   \institute{Departamento de F\'{i}sica de la Tierra y Astrof\'{i}sica, Universidad Complutense de Madrid, Pl. Ciencias, 1, Madrid, 28040, Madrid, Spain
   \and
   Instituto de Física de Partículas y del Cosmos, Universidad Complutense de Madrid, Pl. Ciencias, 1, Madrid, 28040, Madrid, Spain
   \and
   Lund Observatory, Division of Astrophysics, Department of Physics, Lund University, Box 43, SE-221 00 Lund, Sweden
   \and
   Instituto de Astronom\'{i}a, Universidad Nacional Aut\'{o}noma de M\'{e}xico, Circuito Exterior s/n, Ciudad de M\'{e}xico, 04510, Ciudad de M\'{e}xico, M\'{e}xico
             }

 
  \abstract
   { Galaxies are expected to accrete pristine gas from their surroundings to sustain their star formation over cosmic timescales. This mechanism is well established in models and simulations, but evidence from observations is mostly indirect. These gas inflows leave distinct traces in the chemical composition of newborn stars and alter the distribution of stellar abundances compared to what would be expected from a closed-box model of chemical evolution.}
   {The goal of this work is to measure the amount of pristine gas that galaxies accrete during their lifetime, using information on the ages and abundances of their stellar populations and a chemical evolution model. We also aim to determine the efficiency of star formation over time.}
   {We derived star formation histories and metallicity histories for a sample of 8523 galaxies from the MaNGA survey. We use the former to predict the evolution of the metallicity in a closed-box scenario, and estimate for each epoch the gas accretion rate required to match these predictions with the measured stellar metallicity.}
   {Using only chemical parameters, we find that the history of gas accretion depends on the mass of galaxies. More massive galaxies accrete more gas and at higher redshifts than less massive galaxies, which accrete their gas over longer periods. We also find that galaxies with a higher star formation rate at $z = 0$ have a more persistent accretion history for a given mass. We characterize the individual accretion histories in terms of two parameters: the total accreted gas mass and the {\large $\tau$}80 of the accretion history, a measure of when most of the accretion occurred. As expected, there is a strong correlation between the integrated star formation history and the total accreted gas mass, such that more massive galaxies accreted more gas during their lifetime. Currently star-forming galaxies lie above this correlation, so they tend to accrete more gas than average. The relationship between {\large $\tau$}80, the current stellar mass, and the current specific star formation rate is split such that star-forming galaxies (as now observed) may be found in a population with persistent gas accretion regardless of their stellar mass. The star formation efficiency shows similar correlations: early-type galaxies and higher-mass galaxies had a higher efficiency in the past, and it declined such that they are less efficient in the present. Our analysis of individual galaxies shows that compactness affects the peak star formation efficiency that galaxies reach, and that the slope of the efficiency history of galaxies with current star formation is flat.}
 {We show throughout the article that we can obtain information about the processes that regulate the chemical composition of the interstellar medium during the lifetime of a galaxy from the properties of stellar populations. Our results support the hypothesis that a steady and substantial supply of pristine gas is required for persistent star formation in galaxies. Once they lose access to this gas supply, star formation comes to a halt.}

   \maketitle
%
\section{Introduction}
In the standard Lambda cold dark matter (LCDM) paradigm, pristine gas inflows are a natural expectation and serve as the foundation for galaxy growth \citep[e.g.,][]{Finlator2008,Schaye2010,Fraternali2012, Lilly2013,Ceverino2016,Molla2016,Rodriguez-Puebla2016}. However, direct observations of these inflows have remained elusive due to the difficulty in detecting accreted gas \citep[e.g.,][]{Sancisi2008,Fraternali2008,SanchezAlmeida2014a,SanchezAlmeida2017,Cimatti2019}. Furthermore, the accretion of gas and its observable consequences, such as higher star formation rates (SFRs) and lower metallicity, occur at different times, making concurrent observations challenging. \cite{Giavalisco2011} detected large amounts of pristine gas associated with a cluster of galaxies undergoing an ongoing infall process. \cite{Martin2012} and \cite{Rubin2012} also found evidence of cold gas around star-forming galaxies using Fe and Mg absorption lines, although the low covering fractions of the gas suggest that there is not enough to fuel their star formation. In the Milky Way, the estimated rate of HI gas accretion derived from detected HI clouds is around 0.1-0.4 M, which is insufficient to account for the current SFR \citep{Putman2012}.

Observational evidence of gas accretion in galaxies is largely indirect. For example, the Schmidt-Kennicutt relation has shown that galaxies at all redshifts cannot maintain their current SFR for more than 0.5-2 Gyr without replenishment by gas \citep[e.g.,][]{Schmidt1959,Kennicutt1983,Kennicutt1998,Colombo2018,Colombo2020,Sanchez2021a, Genzel2010,Daddi2010,Tacconi2013, Kennicutt1983,Dekel2006,Dekel2009,Fraternali2012, Lilly2013,Davis2016}. Additionally, the gas fraction decreases with redshift at a much lower rate than the stellar density \citep{Prochaska2005,Rao2006,Lah2007}.

The study of chemical abundances also provides indirect evidence of gas accretion, as gas accreted from the cosmic web is believed to be mostly unenriched by metals \citep{vandeVoort2012}. This could explain the low gas metallicities in some star-forming regions in local and high-redshift galaxies \citep[e.g.,][]{Bresolin2012, Ceverino2016} or the strong gas-phase metallicity gradients observed in some high-redshift galaxies \citep[e.g.,][]{Cresci2010}.

Models of chemical evolution in the Milky Way have shown that the narrow metallicity distribution of long-lived stars in the solar neighborhood can only be reproduced with a continuous inflow of relatively low-metallicity gas \citep{Larson1972,Fenner2003,Chiappini2009}, known as the G-dwarf problem \citep[][]{Searle1972,Tinsley1980,Nordstrom2004,Caimmi2008}. These studies predict an exponential decrease in the infall rate with time, with a current value of 0.4 M$_{\odot}$ yr$^{-1}$. Although such predictions are useful for constraining both cosmological models and subgrid recipes used in numerical simulations, it is important to note that the Milky Way is just one galaxy, and differences in gas accretion rates as a function of mass, morphology, and the environment may exist.

For this study, we adopted a similar approach to estimate the gas accretion rates over time for a sample of 8523 galaxies from the MANGA survey. Specifically, we used the star formation histories and stellar age-metallicity relation to calculate the mass of metal-poor gas required over time to dilute the amount of metals predicted by a chemical evolution model \citep[e.g.,][]{Roca-Fabrega2021, Lilly2013}. We explored the differences between galaxies as a function of various parameters.

The paper is structured into the following sections: Sec.~\ref{sec:data} describes the sample and the observational data employed for this study, while Sec.~\ref{sec:analysis} explains the methodology used to derive the star formation and chemical evolution histories (SFHs and ChEHs hereafter), as well as measuring gas accretion. Sec.~\ref{sec:results} presents the results on the gas accretion and star formation efficiency (SFE) histories, as well as the trends with stellar mass, morphology, and the current star formation. Finally, Sec.~\ref{sec:discussion} discusses the results and outlines possible improvements, while Sec.~\ref{sec:conclusions} provides a summary of the findings.

\section{Data}\label{sec:data}
The MaNGA survey \citep[][]{Bundy2015} consists of integral field unit (IFU) spectroscopic observations of a luminosity-selected sample of 10$^4$ local galaxies ($\langle \mathrm{z} \rangle \sim 0.03$).
The spectra were taken at the 2.5 m Sloan Telescope at Apache Point Observatory \citep{Gunn2006} with the BOSS spectrographs \citep[][]{Smee2013} with fiber bundles of different sizes depending on the galaxy \citep{Drory2015}. Additional fibers and fiber bundles are used for flux calibration and sky subtraction \citep{Yan2016}.

The observations were reduced and calibrated using the Data Reduction Pipeline \citep[DRP,][]{Law2016} after which the data cubes are produced, which are the central distributed data product of the project. These have a spatial resolution of about 2.5"/FWHM and a typical spectral resolution of R $\sim$ 2000 over a wavelength range between 3600 \ AA {} and 10300 \ AA {}.

\begin{figure}%
\centering
\includegraphics[width=\columnwidth]{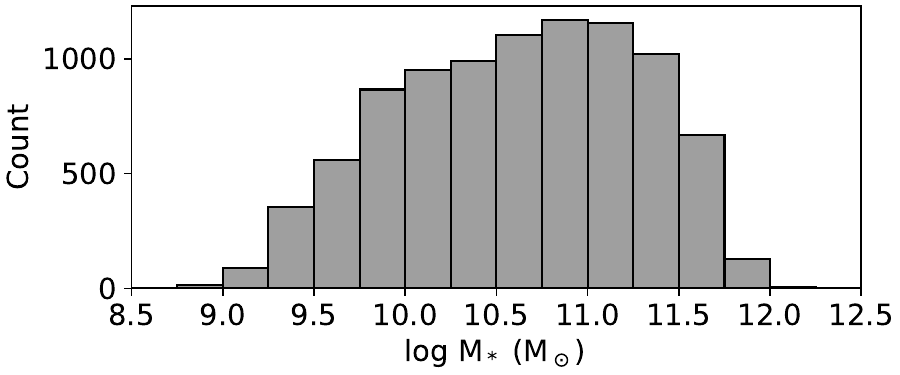}
\caption{Distribution of the sample in stellar mass.}\label{fig:mdist}
\end{figure}

The full sample in the final data release DR17 \citep{Abdurro'uf2022} consists of over 10,000 galaxies, which we refined to the working sample by removing those that showed a poor spectrophotometric fit in the quality control of the reduction and analysis. The procedure to identify galaxies with poor fitting is described in detail in sec. 4.5 of \citep{Sanchez2022}, the first step is an automatic procedure to flag galaxies with anomalous determinations of parameters such as redshift or mass compared to the NASA-Sloan Atlas (NSA)\footnote{\url{http://nsatlas.org/}} catalog. This is followed by human inspection of the central spectrum of each galaxy and its fitted model as well as maps of mock photometry, line emission, ages and metallicities, diagnostic diagrams such as BPT \citep{Baldwin1981} and WHAN \citep{CidFernandes2011}, and the kinematic properties. The ChEH and mass assembly history (MAH) measured at different galactocentric radii are also checked. Galaxies which fail on the inspection are flagged and do not pass the quality check.

Following \cite{Camps-Farina2022}, we also removed galaxies with an inclination greater than 70º or those whose line emission is characteristic of the presence of an AGN, resulting in a working sample of 9087 galaxies. We measure the metallicity at the effective radius, which is difficult to determine in highly inclined galaxies, while AGN can have wide emission lines that interfere when fitting the stellar component of the spectra.
The sample covers a stellar mass range between log M $\sim 8.5 - 12$ M$_\odot$ and includes all morphological types. In Fig. \ref{fig:mdist} we show how the galaxies in the sample are distributed according to their stellar mass.

\section{Analysis}\label{sec:analysis}
\subsection{pyPipe3D}
We used the SFHs and ChEH derived in \citep[][]{Camps-Farina2022} using the methodology of \cite{Camps-Farina2021b}. The metallicity value for each age is measured at the effective radius after averaging the values azimuthally in the projected plane of the galaxy. This value has been shown to serve as a robust proxy for the global properties of a galaxy \citep[][]{GonzalezDelgado2014,Sanchez2020}. Below we give a brief description of the method, but refer to \citep[][]{Camps-Farina2022} for more details. 

SFHs and ChEHs were derived with pyPipe3D \citep[][]{Sanchez2006,Sanchez2016a,Sanchez2016b,Lacerda2022} using the MaStar-sLOG stellar population template library \citep{Sanchez2022}. PyPipe3D is a full spectrum fitting code that can handle both the emission and absorption components of the galaxy spectra and analyze them separately.
The stellar component is analyzed with spectral fitting techniques that obtain the linear combination of simple stellar population (SSP) templates that best reproduce the observed spectrum. The coefficients represent the light fraction contribution of each SSP and can be used to obtain the SFH and ChEH. The fitting procedure is nonparametric, that is, we do not impose priors such as a functional shape of the SFH, which is measured based only on the amount of stellar mass that is assigned to each age. The advantage of this methodology is that we are not limiting the shape of the measured histories but on the other hand we have more free parameters than with a 
parametric approach.

The emission lines of the spectra are fitted by Gaussian functions after the stellar continuum and absorption spectra are subtracted, which allows us to correct their emission for the absorption of the stellar component.
The redshift, velocity dispersion and dust attenuation of the spectra are determined first by performing a fit with a limited set of SSP templates and are then used as inputs when performing the second fitting with the full set of templates.
The redshift and velocity dispersion are used to shift and broaden the templates before using them to fit the spectra.
This two-step procedure has been shown to improve the quality of the spectral fitting \citep[e.g.,][]{Sanchez-Blazquez2011,Sanchez2016a,Sanchez2016b,Lacerda2022}, mitigating existing degeneracies between line-broadening and metallicity.

Light fractions are converted into mass fractions using predicted mass to light ratio (M/L) values. The mass fractions can be used to derive the SFH using the fraction of stellar mass that is lost during stellar evolution after each burst of star formation. The ChEH is measured by averaging the metallicity of the populations measured at each age weighed by the corresponding fractions of light. The metallicity in stellar atmospheres reflects the metallicity of the gas clouds in which they originated. By quantifying the metallicity of stellar populations of different ages, we can obtain a measure of the metallicity evolution within the ISM.

The MaNGA Pipe3D VAC catalog\footnote{\url{https://www.sdss4.org/dr17/manga/manga-data/manga-pipe3d-value-added-catalog/}} contains a large number of parameters of the MaNGA sample derived using PyPipe3D analysis.
In this article, we used the following parameters from this catalog: Stellar mass, current SFR from H$\alpha$ emission, EW$_\mathrm{H_\alpha}$, and the effective radius ($\mathrm{R_e}$). The first two parameters are used to calculate the specific star formation rate (sSFR), and the EW$_\mathrm{H_\alpha}$ is used to determine the star formation status (SFS) of the galaxies according to the prescription of \cite{Lacerda2020}. Star forming galaxies (SFG) are defined as those with EW$_{\mathrm{H}\alpha}$ (R$_\mathrm{e}$) $>$ 10\AA{} and retired galaxies (RG) are defined as those with EW$_{\mathrm{H}\alpha}$ (R$_\mathrm{e}$) $<$ 3\AA{}, whereas Green Valley galaxies (GVG) are defined as those with EW$_{\mathrm{H}\alpha}$ (R$_\mathrm{e}$) between the two aforementioned values.

Morphology was determined using a machine learning algorithm \citep[see][]{Sanchez2022} trained with 6000 galaxies with visual classification by \citep[][]{Vazquez-Mata2022}. We classified the galaxies into E, S0, Sa, Sb, Sc and Sd-Irr morphological bins following the same definition as in \cite{Camps-Farina2022}.

\subsection{Galaxy chemical evolution model}\label{sec:model}

\begin{figure}%
\centering
\includegraphics[width=\columnwidth]{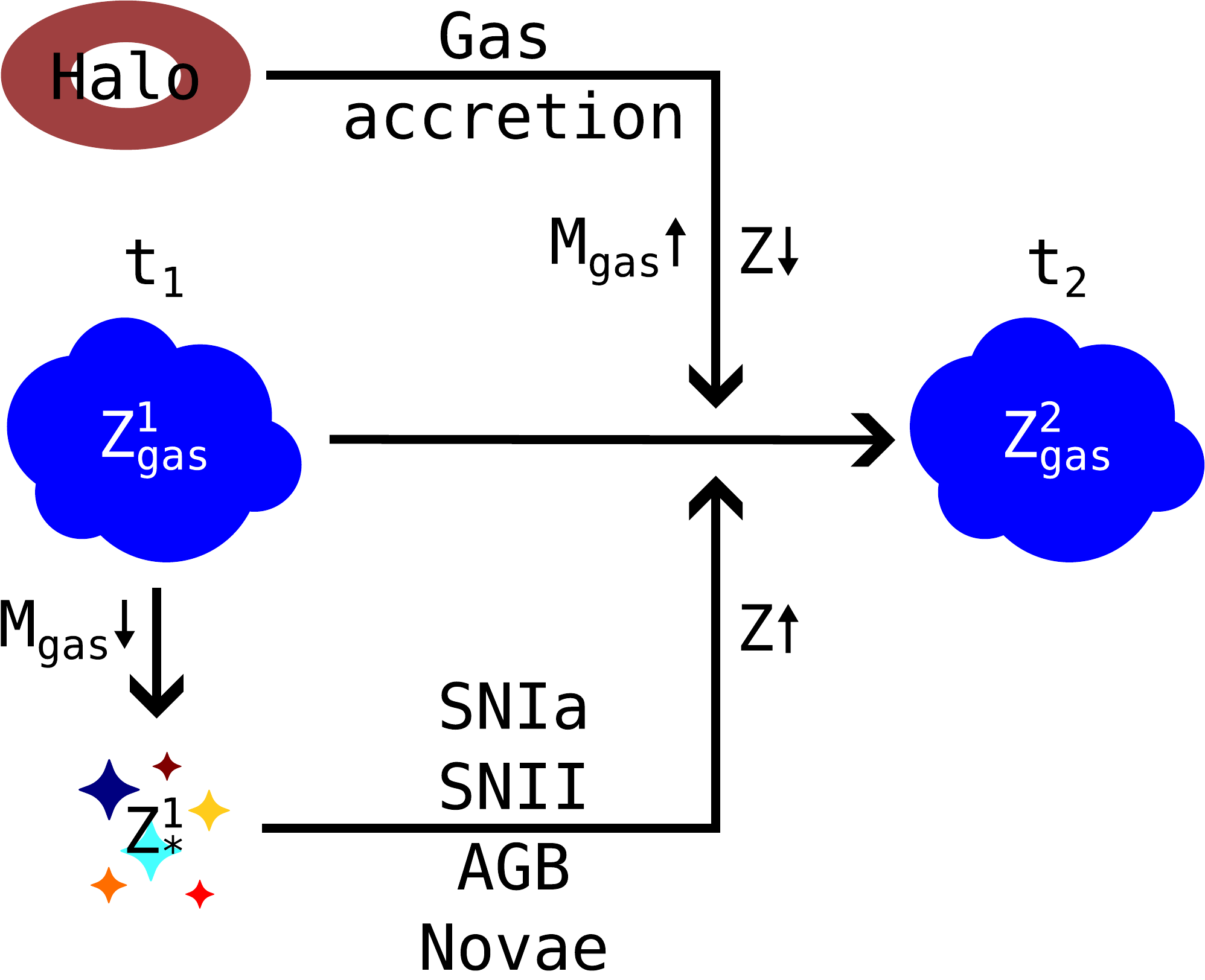}
\caption{Diagram of the model considered for the relation between the ChEH, the SFH and the gas accretion history. The change in the metallicity of the gas over a time interval (t1 to t2) is the result of the metal input from the stellar populations created by the SFH up to that point and the dilution due to the accreted pristine gas. The model does not consider other mechanisms such as outflows or mergers whose effects of dilution or over-enrichment will therefore be conflated with the balance of gas accretion.}\label{fig:model}
\end{figure}

We obtained the evolution of the metal content in a galaxy for a given SFH using the chemical evolution code presented in \cite{Roca-Fabrega2021}.
The code's inputs, aside from the SFH, are the gas accretion history, initial gas mass, and an initial mass function (IMF). The code follows the evolution of the gas-phase chemical abundances of different elements, including Fe-peak and the alpha elements, using the yields by \citep{Kobayashi2011,Stockinger2020} for SNII, \citep{Chiappini1997,Greggio2010,Hillman2015} for SNIa, \citep{Hernanz2005,Izzo2015} for novae and \citep{Ventura2013,Ventura2020} for AGB stars.

We have modified the code to input the SFHs and stellar metallicities to derive the amount of low metallicity gas accretion needed to reproduce the latter. We assume that the accreted gas has an iron abundance of [Fe/H] = $-$1.5, a value based on measurements of damped Lyman-$\alpha$ absorbers at z$\sim$5 \citep{Poudel2020}. We chose a Salpeter IMF \citep{Salpeter1955} for consistency with the SSP templates. The impact of the IMF on the results is mentioned in Sec.~\ref{sec:disc-valid}. The metallicity value of the ISM is initialized to the stellar metallicity derived for the oldest age bin we considered ($\sim12.7$ Gyr), which tends to be substantially higher than the primordial values as stellar population analysis cannot resolve the SFH at its initial burst due to a loss of age resolution among other reasons (see Sec.~\ref{sec:disc-caveats}). 

In order to study the robustness of using the oldest determination of the abundance as the initial value is we can compare our results to the abundance of the oldest populations detected in the Milky Way. In figure 8 of \cite{Minchev2018} the average [Fe/H] at $\sim$12.5 Gyr is -0.45, while averaging the abundance for the oldest age for galaxies with log M$_\star$ = 10-11 M$_\odot$ gives a value of -0.32. This is not too large of a discrepancy even before we take into account that \cite{Minchev2018} only measures stars at or farther than 3 kpc, which is higher than the value of R$_\mathrm{e}$ for the Milky Way at $\sim$2.5 kpc \citep{VandenBergh1999}. The effect of measuring only stars farther than the R$_\mathrm{e}$ would lower the abundance compared to our measurements at R$_\mathrm{e}$ and even so there is no guarantee that the Milky Way should precisely match the average value of galaxies of this stellar mass range. As such, we consider that initializing the ISM at the oldest value is valid and produces reasonable accretion histories.

The initial gas mass is assumed to match the stellar mass at the initial time, an approximation based on a study of the HI content at high redshift by which M$_\mathrm{HI} \sim \mathrm{M_{*}}$ at the redshift values we can reasonably resolve (z$\sim$2-3) \citep{Heintz2022}. In Fig.~\ref{fig:model} we show a schematic representation of the considered model.

The matching of iron abundance with the history of chemical enrichment obtained with spectral population fitting is due to the fact that the metallicity values of the templates used in stellar population synthesis are primarily related to iron abundance, which has the greatest influence on stellar spectra, and therefore is how the templates are labeled \citep[e.g.,][]{Sanchez2021b}.

We have a set of input parameters for each galaxy in our sample, as well as those resulting from averaging the SFH and ChEH of each galaxy in stellar mass, morphology, and the current SFR bins. The averaging procedure was developed to prevent the nonuniform redshift coverage in the MaNGA sample from affecting the averaged histories. The full procedure can be found in \cite{Camps-Farina2022}. The averaged SFHs and ChEHs are more reliable due to their higher statistical significance and allow for a more accurate determination of accretion rates representative of galaxies within a given group. The individual histories, on the other hand, allow us to identify more detailed trends. In Sec.~\ref{sec:results} we show the gas accretion histories resulting from both types of input. What we show as "averaged gas accretion histories" corresponds to the gas accretion that results from using averaged ChEHs and SFHs, rather than averaging the individual gas accretion histories.

We use {\large $\tau$}80, defined as the time (in Gyr) required for a galaxy to accrete 80\% of its gas, to parameterize the shape of the gas accretion history. Low values indicate that galaxies accreted most of their gas at an early time, while high values represent persistent gas accretion over cosmic times.

\begin{figure}%
\centering
\includegraphics[width=\columnwidth]{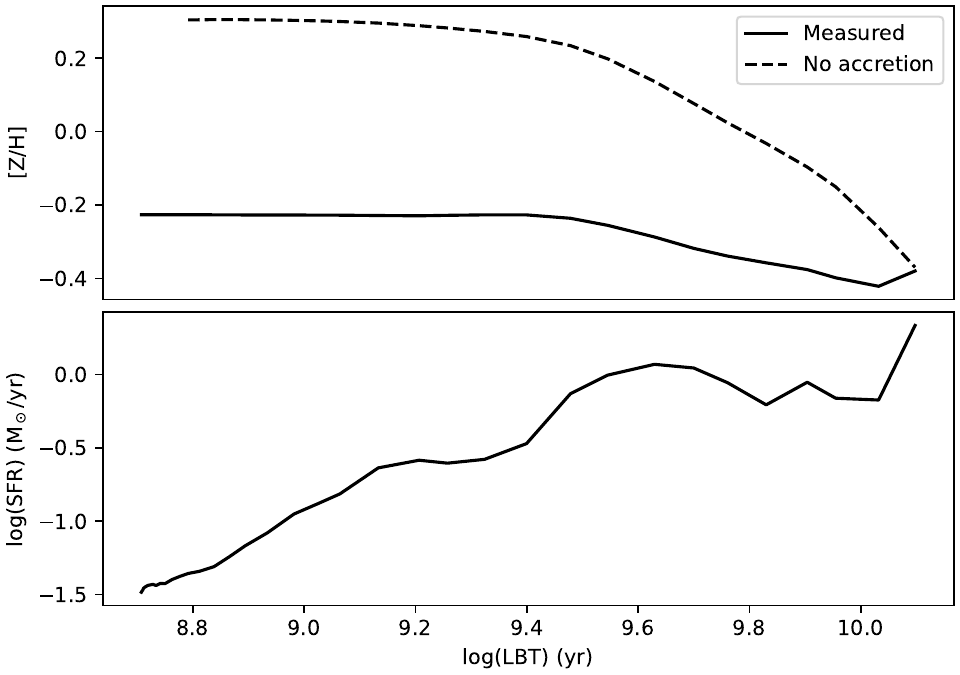}
\caption{Set of ChEH (top) and SFH (bottom) corresponding to the averaged histories of all galaxies in our sample. In the top panel, the solid line corresponds to out measured ChEH and the dashed line to the predicted enrichment history in the absence of dilution from pristine gas accretion.}\label{fig:example}
\end{figure}

As an example of the histories we used as input, and to show the importance of having a mechanism for metal dilution, we show in Fig.~\ref{fig:example} the ChEH and SFH resulting from averaging all the individual histories of the galaxies in our sample, as well as the predicted ChEH without dilution. The amount of metals in the gas increases steadily with time until it reaches a plateau, while the SFR decreases in general.

Some galaxies (about 5\% of the sample) clearly show unphysical gas accretion rate histories and were therefore removed from the sample. The affected galaxies are generally those with low stellar mass and bright emission lines. Lower mass galaxies tend to have poorer signal-to-noise ratios, and bright emission lines (especially if they are broad) can interfere with the spectral fitting procedure that produces our ChEHs and SFHs. This affected $\sim$5\% of the objects, leaving 8523 objects in the sample.

\section{Results} \label{sec:results}
\subsection{Averaged gas accretion histories}

\begin{figure*}%
\centering
\includegraphics[width=\textwidth]{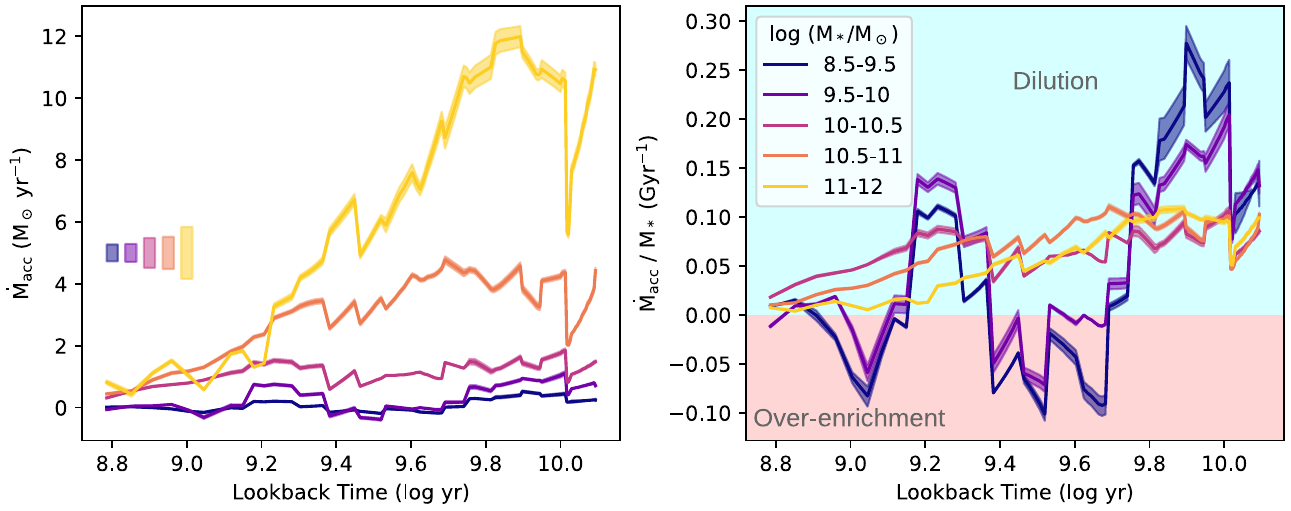}
\caption{Absolute ($\mathrm{\dot M_{acc}}$) and relative ($\mathrm{\dot M_{acc}/M_\star}$) accretion rate histories (left and right panel respectively) as a function of the look-back time for the galaxies in our sample segregated by their current stellar mass (colors). In both panels the shaded areas correspond to the uncertainties derived from the error of the average value in the ChEHs and SFHs. This is not representative of the scatter of the distribution within each bin, which is indicated in the left side of the left panel.}\label{fig:all}
\end{figure*}

In Fig.~\ref{fig:all} we show the average accretion histories in different stellar mass ranges. For each mass bin, we estimated an uncertainty in the mean accretion history by propagating the errors in the mean SFH and ChEH, and we estimated the scatter by propagating the standard deviation of each ChEH and SFH.

It can be seen that the total gas accreted by galaxies throughout cosmic time increases with stellar mass but is very similar when this value is normalized to the stellar mass at $z = 0$. The first statement is obviously expected, but the novelty is that we have only used the difference between stellar metallicity measured from the spectra and that expected in a closed-box chemical evolution model to derive this result, without adding any mass-related constrain. Furthermore, while more massive galaxies accrete more gas relative to their mass at early times, this trend reverses at $z = 0$.

In the lowest mass bins, we also find epochs where the measured stellar metallicity is higher than predicted by the chemical evolution code, which results in a negative value for the accretion rate. These can be simply due to imprecise measurements, especially as they appear to occur for the mass bins with lower accretion rate, but they can also indicate a larger importance of outflows in low mass galaxies, where more metal poor gas is selectively lost. A loss of metal-poor gas results in the average metallicity of the galaxy rising compared to a closed box evolution as the higher metallicity gas remains instead.

\begin{figure*}%
\centering
\includegraphics[width=\textwidth]{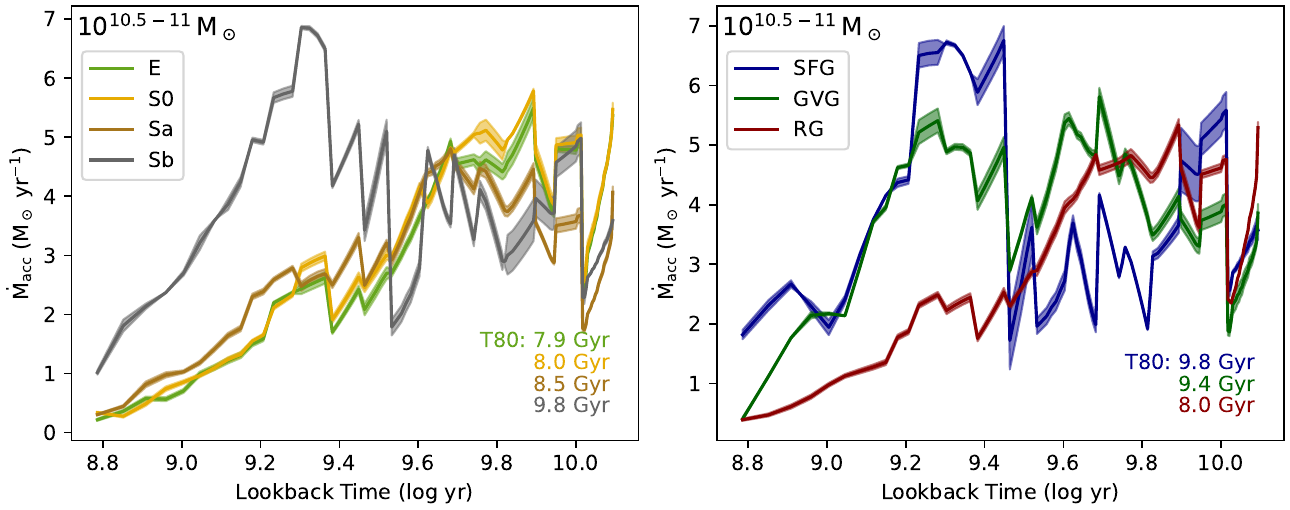}
\caption{Accretion rate histories for different morphology (left panel) and SFS (right panel) bins at a fixed mass bin of $10^{10.5-11}$ M$_\odot$. 
In the bottom right corner of each panel, we show the {\large $\tau$}80 of the accretion histories, defined in Sec.~\ref{sec:model}. In both panels the shaded areas correspond to the uncertainties derived from the error of the average value in the ChEHs and SFHs.}\label{fig:compare}
\end{figure*}

In Fig.~\ref{fig:compare} we show the average mass accretion rates of galaxies in the intermediate stellar mass range $10^{10.5-11}$ M$_\odot$, separated in bins of morphological type and SFS.
It can be seen that the accretion rate for all morphological types increases at early times, reaches a maximum, and then decrease almost continuously, reaching their minimum values at $z = 0$. The decrease starts at 7, 4 and 2 Gyr ago for E-S0, Sa and Sb types respectively, which is reflected in the different {\large $\tau$}80 values. This progression should be at least partially related to the fact that later morphological types are more likely to be star-forming and, indeed, it can be seen in the figure that SFG and GVG have an accretion rate history very similar to Sb galaxies, while the evolution of the mass accretion rate in RG follows that of E and S0s.

The similarity between the accretion histories of SFG and GVG is particularly interesting. The most recent peak of accretion at $\sim10^{9.3}$ yr ($\sim$2 Gyr) is lower for GVG but otherwise both bins have had accretion episodes throughout cosmic time. 1 Gyr ago, the accretion rate for GVG fell to the values of RG separating from SFG.
This similarity is in very good agreement with what we would expect as GVG are galaxies that are just now becoming retired, meaning that their accretion history up until this point should be similar to that of SFG, as is observed.

\cite{Keres2005} and \cite{vandeVoort2012} use hydrodynamic cosmological simulations to estimate the gas accretion rates of galaxies by tracking the kinematics of the gas around them from the cosmic web. For Milky Way-like halos they predict accretion rates which peak at around 10-30 M$_\odot$ yr$^{-1}$ at z$\sim$ 2-3 and drop to about 1-3 M$_\odot$ yr$^{-1}$ in the present. For similar stellar masses we obtained an average accretion rate a factor of 2-6 lower at the peak ($\sim$ 4.8 M$_\odot$ yr$^{-1}$) at the peak, but a very good agreement in the local universe (1.2 M$_\odot$ yr$^{-1}$).
Given the agreement on the recent value it is possible that the discrepancy in the peak value is due to us not being able to observe the peak of the accretion within the LBT range that we can reliably resolve. Alternatively, the simulations might overestimate the early accretion rate as a result of the subgrid physics recipes used or due to resolution issues.
The choice of 1 Gyr ago for the recent value instead of the most recent measurement is due the fact that the accretion rates measured for the most recent times (relative to when the light was emitted) are likely to be underestimated simply due to the delay between the infall of the gas, it becoming well-mixed to the ISM and finally stars being born whose metallicity has been affected by the accretion.

Stellar population analyzes are especially suited for drawing conclusions by measuring the quantities in relative terms rather than for absolute values, such as findings on which objects or areas of a galaxy are younger or have higher abundances than others. The reason for this is that the values of the properties can change depending on the templates used for the fitting \citep[e.g., see][]{CidFernandes2014}. For example, \cite{Muzzin2009} uses three stellar population models to perform SED fitting, finding a 25-50\% variation for the stellar mass and SFR. Combined with the substantial simplifications in the model (whose impact is discussed in Sec. \ref{sec:disc-valid}) we consider the discrepancy with the aforementioned simulations to be fairly reasonable, especially as the difference between the two hydro-dynamical estimates of the accretion rate themselves is a factor of 3, which is higher than the factor between our results and the closest of the estimates. Additionally, we do not have independent measurements of the abundance of the primordial gas that is accreted, and we assume it does not change its abundance over time. If the accreted gas has a different abundance than the one we assume, this would conversely apply a scaling factor to the accretion rates we measured. Future work on the chemical evolution model should improve the accuracy of the estimates. 

Another important point to consider is the fact that our measurements are akin to an "effective" gas accretion rate. As stated above, in order for the accreted gas to be measured using our method it not only has to be accreted into the galaxy, but it must also reach the loci of future star formation before a star formation burst is triggered. The aforementioned studies measure the accretion rate of the gas from its kinematic infall, not considering how much of this gas will eventually condense into H$_2$ and form stars. Galactic winds with high mass-loading factors in particular can also remove a substantial portion of the accreted gas before it has a chance to form stars \citep[e.g.,][]{Veilleux2005,Shen2012,Genzel2014,Lopez-Coba2019,Lopez-Coba2020,Concas2022}. This removal is dependent on the mass of the galaxy and is thus not simply a general scaling factor in our results.

\subsection{Star formation efficiency} \label{sec:sfe_aver}
\begin{figure*}%
\centering
\includegraphics[width=\textwidth]{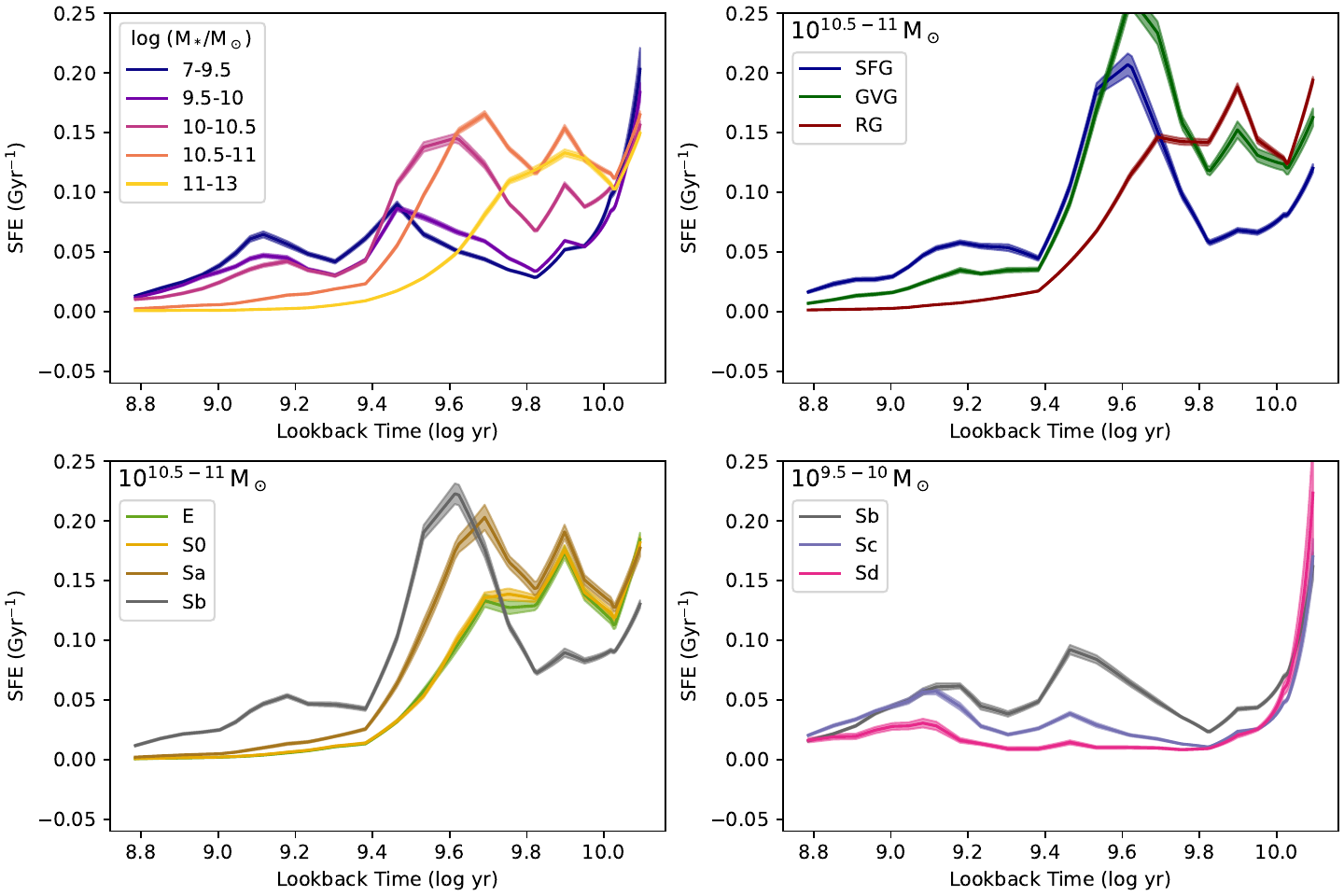}
\caption{SFE histories for different bins of galaxies. In each panel the X-axis is the look-back time in log and the Y-axis the SFR divided by the total amount of gas present at each time as predicted by the model. In the top left panel, we show the galaxies divided into stellar mass bins, in the top right panel we show one mass bin (10$^{10.5-11}$ M$_\odot$) divided into current SFS bins, in the bottom left panel we show the same mass bin divided into four morphology bins (E-Sb) and in the bottom right panel we show a different mass bin (10$^{10.5-11}$ M$_\odot$) divided into the Sb, Sc and Sd morphology bins.}\label{fig:sfe_comb}
\end{figure*}

Given that we now have the accretion histories as well as the SFHs of the galaxies in each bin, it is trivial to compute the SFE histories (SFEHs) as well. Unlike the accretion rate, the SFE is inversely proportional to the total gas mass present in the galaxy at each time. As a result, it is more affected by the choice of the initial gas mass. In Appendix \ref{sec:appendix} we show how changing the value of the initial gas mass has a much higher effect on the SFEH than on the accretion rate history.

Another important point to consider is which phase of the ISM the gas mass we are measuring with our gas mass parameter. The gas accretion we measured is derived from matching the abundance of the ISM to the stellar populations, so whether it is HI, HII or H$_2$ depends on which phase the mixing occurs. Given that the evidence for accretion of gas is observed in HI and that the dense clouds that form stars are made up of H$_2$ we generally expect our measurement of the gas mass to refer primarily to HI or to the total sum of gas. It is also important to keep in mind that even if the gas is accreted and mixed in the HI phase this does not guarantee that several Gyr after accretion it will still be in HI form, the fact that stars are formed out of it (a prerequisite to be measured in our methodology) means that at least a portion of it was converted into H$_2$. We removed the amount of stellar mass formed from the amount of gas, but this is only a lower bound to the amount of gas converted into H$_2$. There is also some amount of gas which will be ionized by the stellar populations into HII, so in general our gas mass estimates correspond to somewhere between the total mass of HI and the total gas mass in a galaxy.

In Fig.~\ref{fig:sfe_comb} we show the averaged SFE of galaxies in bins of stellar mass, SFR and morphology. The two morphology panels (bottom ones) use two different mass bins so that the late-type galaxies can also be shown while maintaining good statistical representation in each bin, with the Sb bin serving as a bridge between the two panels.

It can be seen that the peak of the SFE occurs at higher redshift for more massive galaxies. It is also clear that the decline of the efficiency after the peak is steeper for massive galaxies. This leads to the current segregation such that less massive galaxies are more efficient at forming stars in the present day. At LBT$\sim10^{9.4}$ yr the trend with stellar mass has inverted and less massive galaxies have higher SFE. The two less massive bins have similar SFE histories though they still show the same trend of more massive galaxies being more efficient in the past and less so in the present.

At a given stellar mass, the SFE decline after the peak depends on the current star formation status. RG dropped earlier in SFE compared to GVG and SFG. After LBT$\sim10^{9.4}$ yr the three bins show a fairly flat histories but SFG consistently maintain a higher SFE compared to GVG and RG.

Regarding morphology similar results can be seen, likely affected by the correlation between morphology and SFS, with the earlier type galaxies in the E-Sb range showing an earlier peak of SFE which declines more than the rest which produces an inversion in the M$_\star$-SFE correlation. The lower mass, late type galaxies in the bottom right panel appear to follow the trend of the less massive galaxies in the top left panel, with similar SFEHs but in this case the earlier type galaxies are more efficient at all times. These galaxies as well as the low mass ones on the top left panel show a divergence at the earliest times toward high SFE values, likely pointing to the fact that our prescription for the initial gas mass underestimates the value for these galaxies and that even at the earliest ages that we can resolve the gas fraction for late-type and lower mass galaxies is higher than for more massive, earlier type galaxies.

One could interpret the fact that the inversion in SFE is barely detected for Sb and Sc and not at all for Sd in the bottom right panel as a difference in timing rather than in the physics involved. For lower mass Sb and Sc galaxies it could be that the drop in their average SFE, observed around LBT$\sim10^{9.3}$ yr for higher mass galaxies, has not yet occurred. Indeed, for Sb galaxies in this panel the most recent values show their SFE dropping slightly below Sc.

In order to compare our values for the SFE with observations we need to consider the gas phases. We proceed under the assumption that the gas is accreted as HI but part of it is converted into H$_2$ and HII according to the ratios which are currently observed. \cite{Calette2018} put the H$_2$ to HI ratio for galaxies with log M$_\star$ $\sim10.5$ M$_\odot$ (the bin we use for the comparison) at about 0.4 dex for late type galaxies, while the ratio of HII to HI is estimated as 0.1 in \cite{Kado-Fong2020}. As such, we can use these ratios to obtain estimations for each gas phase in terms of our measured gas mass as:
\begin{itemize}
  \item M$_\mathrm{HI}$ $\sim$ 0.67 M$_\mathrm{gas}$
  \item M$_\mathrm{H_2}$ $\sim$ 0.26 M$_\mathrm{gas}$
\end{itemize}
Our measured SFE for SFG galaxies of $10^{10-10.5}$ and $10^{10.5-11}$ M$_\odot$, is $\sim$0.05 Gyr$^{-1}$ 1 Gyr prior to observation, which can be converted under our assumptions to:
\begin{itemize}
  \item SFE$_\mathrm{HI}$ $\sim$ 0.08 Gyr$^{-1}$ ; $\tau_\mathrm{dep} \sim 12.5$ Gyr
  \item SFE$_\mathrm{H_2}$ $\sim$ 0.19 Gyr$^{-1}$ ; $\tau_\mathrm{dep} \sim 5.2$ Gyr
\end{itemize}

\cite{Leroy2008} find a value of 0.5 Gyr$^{-1}$ averaging measurements of the SFE in local galaxies of log M$_\star$ $\sim 10.1-10.9$ M$_\odot$ in the SINGS survey using molecular gas and \cite{Colombo2018} similarly computes the molecular gas depletion time for EDGE-CALIFA galaxies finding a typical value of $\sim10^{9.5}$ yr (except for E galaxies) which corresponds to $\sim 0.3$ Gyr$^{-1}$.
Regarding studies on SFE$_\mathrm{HI}$, \cite{Parkash2018} and \cite{Chowdhury2022} find depletion times of 3-6 Gyr which correspond to 0.15-0.33 Gyr$^{-1}$.
For both phases our estimations of the SFE are a factor of 2-6 times lower than observational values which is expected given that we do not consider the effect of outflows on the amount of gas within the galaxy. Outflows could be responsible for expelling substantial amounts of gas, thus lowering the remaining mass causing us to underestimate the SFE, especially at later times. A removal of around 20-80\% of the total amount of accreted gas via outflows over the lifetime of these galaxies would give a reasonable match with the results and this is actually a relatively low value compared to the expected mass loading factors in galaxies from simulations \citep{Mitchell2020} and observations \citep{Chisholm2017} with $\mathrm{\dot M_{out} / \dot M} \geq 1$.

\subsection{Individual gas accretion histories}\label{sec:indiv}
From this point on we used the individual accretion histories of the 8523 galaxies in the cleaned sample instead of the averages used in previous sections.
As we are dealing with many objects, instead of the accretion histories, we focus on two quantities that parameterize them: (i) the total accreted gas mass and (ii) the {\large $\tau$}80 of the accretion history. Negative values of the accretion rate were set to zero prior to calculating these two parameters. These negative values appear sometimes in the accretion histories and can arise due to method inaccuracies, or they could be due to effects that produce an over-enrichment of the ISM such as outflows of low-metallicity gas (thus raising the average metallicity).

The code tries to match the abundance of the gas by adding material of lower abundance and therefore the main assumption that is expected to hold true is that the measured abundances are equal or lower than if no gas was accreted. A higher value of the abundance than expected would therefore give a negative numerical value to match the observed value. As with any measurement, there are errors associated with the determination of the abundances and it is therefore expected that in some cases the aforementioned main assumption fails to hold true purely due to uncertainty fluctuations. This is more likely for lower mass (and consequently brightness) galaxies which also tend to have lower abundances in general, thus increasing the errors and lowering the values at the same time. Stellar population synthesis is also affected by an intrinsic degeneracy between age and metallicity such that an increase in age and in metallicity produce similar changes in the template spectra. The effects it has on the accretion rate histories are discussed in Sec. \ref{sec:disc-caveats}.
While there is significant value in checking for the existence of galaxies whose over-enrichment is physical in nature, for the purposes of this article we are focusing on dilution effects and negative values make the measurement of two aforementioned parameters ambiguous.

\begin{figure*}%
\centering
\includegraphics[width=\textwidth]{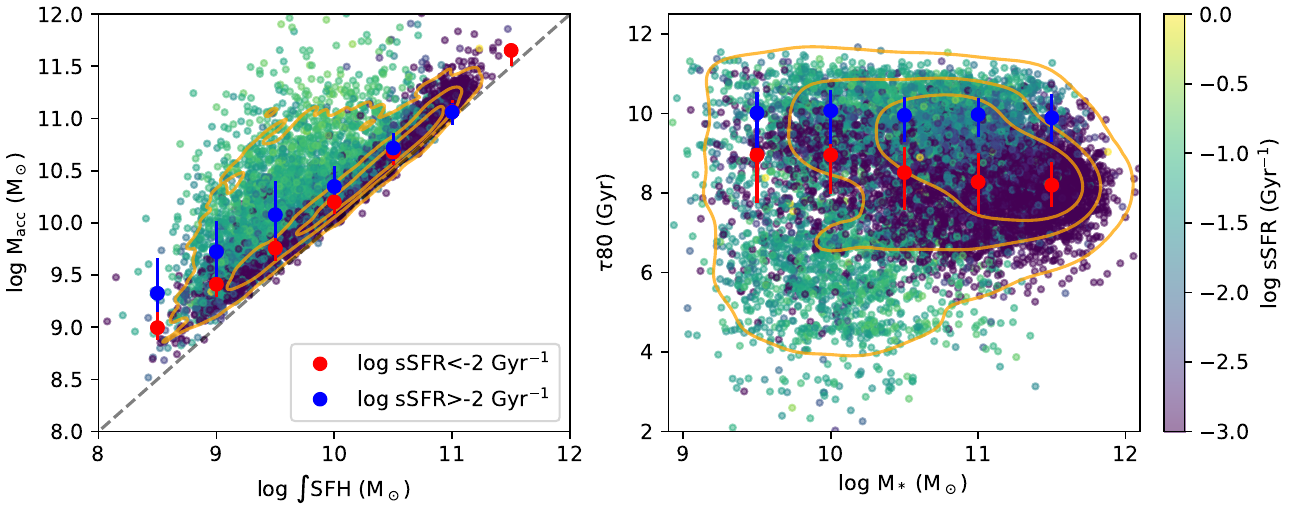}
\caption{Results for the gas accretion histories obtained using the chemical and star formation history of each galaxy. In the left panel, we show the relation between the integrated SFHs and the accreted gas masses and in the right panel we show the relation between the current stellar masses and the {\large $\tau$}80s of the accretion histories. In both panels the color corresponds to the current sSFR values and the contours enclose 35\%, 65\% and 95\% of the distribution. The thick points correspond to the median values within 0.5 dex wide mass bins separated into two sSFR ranges with the error bars marking the 25th and 75th percentiles. In the right panel, the log sSFR > -2 Gyr$^{-1}$ averaged values do not include the cloud of points located below 8 Gyr in {\large $\tau$}80.}\label{fig:mass_t80}
\end{figure*}

In the left panel of Fig.~\ref{fig:mass_t80}, we show the relation between the total amount of stellar mass formed in a galaxy ($\int$SFH), the total mass of the gas it accreted, and its current sSFR. It is important to note that due to the fact that we do not resolve the earliest times in the SFH combined with mass loss due to aging stellar populations, the integrated SFH is not the same value as the current stellar mass. The former is the result of integrating the SFH derived from the stellar populations while the latter is measured from its current luminosity and applying the appropriate M/L in the V band \citep[see][]{Sanchez2022}.

There is a clear, very tight, correlation between the amount of stellar mass a galaxy produces and the amount of gas it needs to accrete in order to match the observed chemical content. There is also a secondary correlation with the sSFR such that galaxies which are currently still forming stars accrete more gas compared to others of similar stellar mass. This result is especially remarkable as the sSFR is determined from emission lines and is therefore only a measurement of the current star formation of galaxies. The correlation provides a clear link between a galaxy's past history of accretion and its current star formation properties.

The gray dashed line shows the 1:1 relation between the two quantities, with practically all galaxies located above the line and the bulk of the distribution lying just above and parallel to it. The absence of galaxies which have accreted less gas than they have consumed to form stars is in very good agreement with the hypothesis that star formation is fueled by gas accretion. The reality of this process, however, is much more complex once we account for the different phases of matter (HI, H$_2$, HII), the infall, mixing and star formation timescales and the effects of outflows on the balance between accreted gas and stellar mass formed. The fact that galaxies lie above rather than centered on the 1:1 line is probably the result of a combination of these factors. For example, the presence of metal-rich outflows would simultaneously dilute the gas in the absence of accretion, making us overestimate the amount of gas in the galaxy, and also remove material available to form stars, moving a galaxy up and left in the relation making it seem like an outlier with very low SFE. 

The fact that star-forming galaxies can lie at an order of magnitude or more above the relation is likely partly explained by feedback removing or heating gas so it is not available for star formation. Another hint to the effect of outflows could be that for lower $\int$SFH galaxies ($\sim10^9$ M$_\odot$) in the left panel of Fig.~\ref{fig:mass_t80} the relation curves slightly upward. In these galaxies outflows are expected to be more efficient due to lower escape velocities so there should be, on average, a larger difference between the amount of gas accreted and the amount that is left for star formation after outflows remove a portion.

The efficiency of the conversion of accreted gas, which is mostly atomic, into molecular gas also plays a role in this, though typical recipes for the conversion cannot be applied directly. Since we measured the accreted gas using the information on the chemistry of the stars, the gas we are measuring has at the very least mixed with the gas that ends up forming the stars, the phase of matter (HI, HII, H$_2$) in which the mixing occurs affects how much of the measured accreted gas mass contributes to fueling star formation in the galaxy and therefore the actual SFE.

In the left panel, the binned data points for the two ranges of sSFR (red and blue dots) show a convergence such that they become more similar for higher stellar masses, which could also be the result of stellar feedback outflows becoming more efficient at lower stellar masses. They also show that the bulk of the galaxies lies close to the 1:1 line and that the galaxies that accrete very high amounts of gas relative to their stellar mass are a minority.

Other than how much gas is accreted over a galaxy's lifetime it is also important to understand how this accretion is distributed over its lifetime. In the right panel of Fig.~\ref{fig:mass_t80}, we show the correlation between the {\large $\tau$}80 parameter of the accretion history and their current stellar mass and sSFR.
The larger cloud of points in the figure shows a correlation between the three parameters such that currently star-forming galaxies lie in a fairly compact sequence of high {\large $\tau$}80 for all stellar mass values whereas retired galaxies appear to be centered at higher masses but also at significantly lower {\large $\tau$}80 values. This distribution is reminiscent of the star formation main sequence, and it shows that galaxies that are currently star-forming have overall a persistent gas accretion or have had a significant recent accretion event, while galaxies whose accretion has declined tend to be retired nowadays.

There is another cloud of points at low mass and low {\large $\tau$}80 which also has high value of the sSFR, suggesting the existence of galaxies which are still star-forming but have not required recent accretion to dilute their gas compared to earlier epochs. Since in the left panel star-forming galaxies consistently accrete more gas without showing a separate cloud of points it is possible that these galaxies are still accreting enough gas to sustain star formation, but they had an unusually high accretion rate in the past, shifting the {\large $\tau$}80 to lower values.

The binned data points show the offset between galaxies as a result of sSFR as well as a lack of correlation of {\large $\tau$}80 with stellar mass beyond a slight decrease at higher masses for low sSFR galaxies. An important thing to note is that the calculation of the high sSFR data points does not include the secondary cloud of low {\large $\tau$}80 points which have been masked prior to their calculation so that they show the trend of what we consider to be the main distribution.

As a whole, Fig.~\ref{fig:mass_t80} shows that for a galaxy to still be forming stars today it needs: (i) higher amounts of accreted gas and (ii) for the accretion to not have stopped or heavily declined in the recent past. This is compelling evidence for the scenario in which persistent star formation in galaxies occurs primarily as a result of ongoing accretion of pristine gas.

\subsubsection{Star formation efficiency}

\begin{figure*}%
\centering
\includegraphics[width=\textwidth]{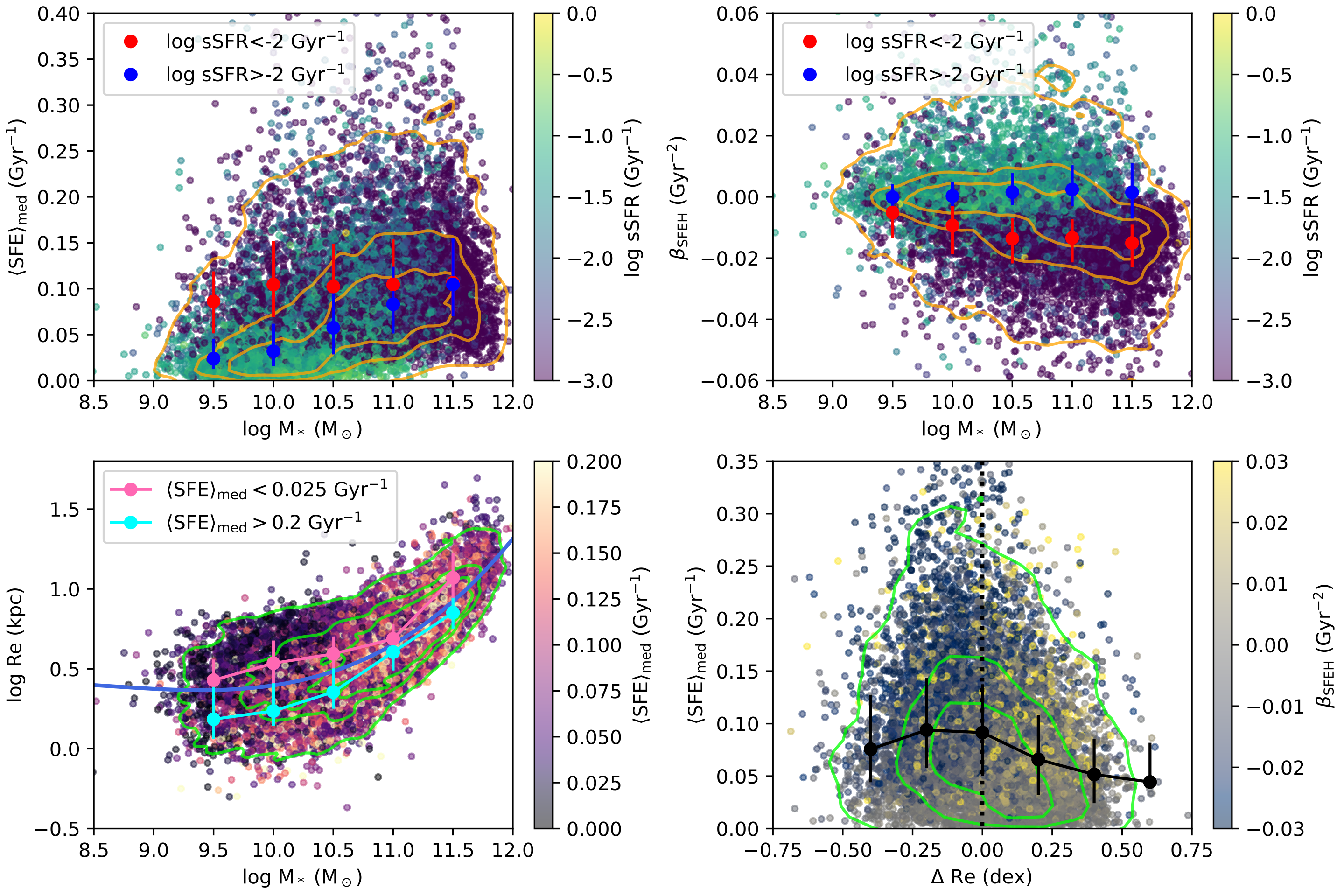}
\caption{Results for the SFE histories obtained using the chemical and SFH of each galaxy. In the upper left panel, we show the relation between stellar masses and median SFEs and in the upper right panel the relation between stellar masses and the slope of the SFE histories. In both of these panels the color corresponds to the sSFRs.
In the bottom left panel, we show the relation between the stellar masses and the effective radii with the median SFEs shown in color. The blue line corresponds to a cubic fit to the M$_\star$-Re relation. In the bottom right panel, we show the relation between the ratio of Re to the value of the cubic fit shown in the left panel and the median SFE, with the slope of the SFE histories shown in color. The contours in each panel enclose 35\%, 65\% and 95\% of the distribution respectively. The thicker points shown in each panel correspond to the median values of the Y-axis quantity within 0.5 dex wide bins in the physical quantity of the X-axis selected as indicated in the legend for each color, with the error bars marking the 25th and 75th percentiles.
}\label{fig:ind_sfe}
\end{figure*}

Much like for the accretion histories, we need to parameterize the individual SFE histories in order to be able to study them. The SFE is an intensive quantity, meaning that it does not scale with volume or mass like the SFR. If we "double" a galaxy by making it have twice as much material of each type, the resulting galaxy will have double the mass and SFR, but the SFE will stay the same because both SFR and gas mass double, canceling each other's increase. Because of this, the parameters used in the previous section to characterize the accretion rate are no longer relevant for the SFE because the cumulative sum of the SFEH does not have a physical meaning. As a result we have characterized the SFE histories using the median SFE value over time, $\langle$SFE$\rangle_\mathrm{med}$, and the slope of a linear fit of the history, $\beta_\mathrm{SFEH}$. In order to avoid the effect of early LBT divergences due to the underestimation of the initial gas mass seen in Fig.~\ref{fig:sfe_comb} we removed the earliest 2 Gyr of each SFE history prior to calculating the parameters.

In the top panels of Fig.~\ref{fig:ind_sfe}, we show the relation between $\langle$SFE$\rangle_\mathrm{med}$, $\beta_\mathrm{SFEH}$ and the current stellar mass and sSFR of the galaxies in our sample. Similarly to the results observed in Fig.~\ref{fig:sfe_comb} we find that higher mass galaxies have, in general, a higher $\langle$SFE$\rangle_\mathrm{med}$ but a steeper decline over cosmic time. The $\langle$SFE$\rangle_\mathrm{med}$ vs M$_\star$ relation has a very large dispersion which increases for higher masses, but average trends can still be discerned, which generally coincide with those found for the averaged histories.

Currently star-forming and retired galaxies appear to have different distributions within the full sample: The most star-forming galaxies appear to cluster at low $\langle$SFE$\rangle_\mathrm{med}$ with no correlation with M$_\star$, but calculating the median $\langle$SFE$\rangle_\mathrm{med}$ within mass bins and the two sSFR groups shows the underlying correlation. While below $10^{10.5}$ M$_\odot$ the trend for star-forming galaxies is fairly flat, similar to what can be seen in the top left panel of Fig.~\ref{fig:sfe_comb}, it steepens for higher masses. For galaxies with lower sSFR the relation is practically flat regardless of stellar mass, producing a convergence between the two sSFR groups at high stellar masses.

The top right panel shows how star-forming galaxies are tightly clustered around a value of 0 for $\beta_\mathrm{SFEH}$ which means that, on average over their lifetimes, they maintain a similar SFE value. Lower sSFR galaxies, on the other hand, have negative slopes on average, indicating a decline in SFE over time. The median slope decreases for higher stellar masses indicating that for these galaxies the decline in SFE is steeper.
The apparent conclusion is that galaxies which used to form stars at a very high efficiency in the past end up becoming retired in current times, which could be evidence for stellar feedback-based quenching mechanisms but also for other mechanisms that are indirectly related to high SFE. Merger-triggered star-bursts, especially followed by an AGN, would also fit the picture as would morphological quenching of early disk galaxies.

The physical size of galaxies is a parameter that is expected to correlate with their SFE such that smaller galaxies are more efficient at forming stars than larger ones \citep[e.g.,][]{Young1999}. In the bottom panels of Fig.~\ref{fig:ind_sfe}, we check whether this applies to our SFE histories. In the left panel, we show the M$_\star$-Re relation colored by the median SFE of the galaxies with a cubic fit between M$_\star$ and Re.
Below $\sim10^{11.5}$ M$_\odot$ the distribution appears to be skewed with $\langle$SFE$\rangle_\mathrm{med}$ such that more efficient galaxies lie below the average Re and conversely less efficient ones lie above. This trend is confirmed by dividing the sample into high and low-efficiency groups and calculating the median within stellar mass bins. Galaxies with very high (low) $\langle$SFE$\rangle_\mathrm{med}$ follow a trend that is parallel to the full sample cubic fit but lies below (above) the fit such that they are smaller (larger) than galaxies of the same stellar mass.

In the bottom right panel, we try to find a more robust determination of this correlation and also to tie it to $\beta_\mathrm{SFEH}$. We calculated the ratio between the values of Re we measured and those predicted by fit for their stellar mass, thus removing the dependence on M$_\star$, and plotted this ratio versus the $\langle$SFE$\rangle_\mathrm{med}$ with the $\beta_\mathrm{SFEH}$ as the color.

Despite the width of the distribution, the binned data points show a weak trend such that galaxies become less efficient the bigger they are relative to their average M$_\star$. It is particularly interesting that the relation plateaus below a ratio of 1 but declines quickly above this value. The leftmost bin does drop again but we might be affected by that bin having less galaxies.

Compared to those shown in \cite{Young1999} our correlations are much weaker, a result that is likely to be affected (other than by the lower precision of our measurements) by the fact that we account for the correlation between the effective radius of the galaxies and their stellar mass but they do not, showing the correlation between radius and SFE. As such the correlation they find might be affected by both parameters correlating separately with stellar mass. We are also considering the median SFE over their lifetimes rather than the current measurements.

\subsubsection{Matching the remaining gas to observations}
\begin{figure}%
\centering
\includegraphics[width=\columnwidth]{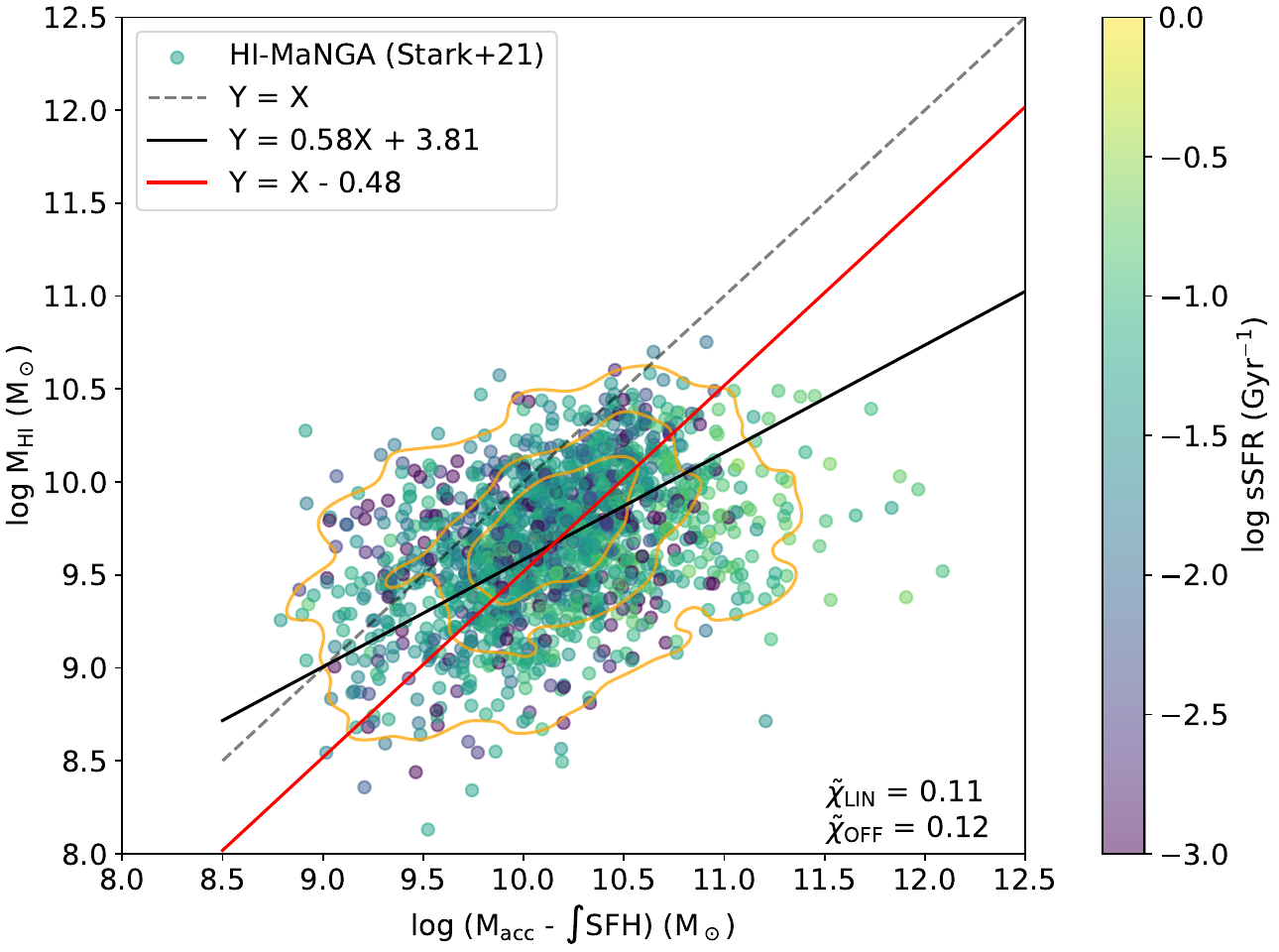}
\caption{Comparison between our predicted remaining amount of gas and measurements of the current HI gas mass from the HI-MaNGA survey \protect\citep{Stark2021}. The dashed line shows the 1:1 relation and the solid lines two different fits to the data, either a linear one or one offset from the 1:1 ratio. The color corresponds to the current sSFR of the galaxies.
}\label{fig:ind_mhi}
\end{figure}
The validity of our measurements of the SFE depends on how accurate our estimation of the remaining gas in the galaxies is, whose main caveat is the fact that we do not consider outflows. As such, we expect our gas mass values to be overestimated by a certain factor, which in turn means that our SFE values are underestimated, as shown in Sec.~\ref{sec:sfe_aver} for the comparisons to literature values.

The HI-MaNGA \citep{Masters2019} survey is a follow-up program for MaNGA which aims to provide information on the neutral gas for galaxies in the MaNGA sample. In the second data release \citep{Stark2021}, they provide HI data for 3818 galaxies of which 1809 have a HI gas mass determination. The intersection with our refined sample is 1371 objects.
In Fig.~\ref{fig:ind_mhi} we compare the predicted current amount of gas, obtained by subtracting the total stellar mass formed from the total accreted gas mass and the initial gas mass, to the measurements of HI gas mass in the galaxies. 
The 1:1 relation is located as an upper envelope of the distribution, mostly parallel to the centermost contour where more objects are present. The galaxies with higher predicted gas mass, which are generally also the most massive in stellar mass, show a significantly lower ratio of measured-to-predicted gas mass, consistent with predictions of the outflow rate due to AGN \citep[e.g.,][]{Mitchell2020}.

We show two fits to the data, one linear and one allowing only an offset to the 1:1 relation. The linear fit was performed using orthogonal distance regression (ODR) to ensure that the slope of the relation follows the overall direction of the distribution, as it minimizes perpendicular distance to the regression rather than distance in the Y-axis. The fitted offset shows on average a 66\% loss of the accreted gas, either by being expelled in outflows or converted into other phases of matter such as molecular and ionized gas.
The linear fit has a shallower slope as a result of the under-prediction of gas mass at low masses and over-prediction at high masses. While the shape of the distribution at the outermost contours favors the linear fit, the central parts appear to follow the offset distribution better, seen in the center-most contour. Overall, the linear fit is better as seen in the reduced chi-square values but the offset one is not much worse.

The over-prediction of gas mass for high stellar mass galaxies fits well with the expectations from simulations. Simulations predict that above log M$_\star \sim 10.5$ M$_\odot$ AGN feedback becomes efficient \citep[e.g.,][]{Dekel2006,Mitchell2020}, which would increase the effect of outflows removing gas. A relative increase in outflow efficiency above these masses would produce the observed widening of the gap between the measured gas mass and the one predicted in the absence of outflows. While the X-axis in Fig. \ref{fig:ind_mhi} does not represent stellar mass directly, more massive galaxies will, on average, have more total gas mass so it serves as a proxy.
The other end of the relation, which shows that low-mass galaxies can have higher measured gas masses compared to our predictions in the absence of outflows, could be the result of underestimating the initial gas mass for these objects. In Sec.~\ref{sec:sfe_aver} we proposed this underestimation as a way to explain the sharp upturn of the SFEH at the earliest ages and it would also explain why we measured lower amounts of gas compared to observations.

Knowing the ratio between the gas masses we predict and those that are measured, we can use the fitted functions to correct the SFE values in our galaxies. In Sec.~\ref{sec:sfe_aver} we compared our results for the log M = 10.5-11 M$_\odot$ bin with literature values, finding that our SFE values are too low in general. Fig. \ref{fig:ind_mhi} shows that we are, in general, overestimating the amount of gas in the galaxies, so after correcting our measured gas masses using the fitted functions we find HI depletion times of 2.9 Gyr (linear) and 4.1 Gyr (offset) while for H$_2$ the depletion times become 1.2 Gyr (linear) and 1.7 Gyr (offset). The literature values for HI are 3-6 Gyr and for H$_2$ it is $\sim$2 Gyr.


Both versions of the SFE now fit very well with the results from the literature but this is to be expected as our "correction" effectively transforms our gas masses into the average values of the HI-MaNGA survey whose gas masses have been validated against other works in the literature.
The main result of this section is the fact that our gas mass estimations are reasonable given the approximations in the model and the fact that the trends, such as more massive galaxies having lost more gas to outflows, make sense given the physics we have implemented in the model.

\section{Discussion}\label{sec:discussion}

The results shown here provide a consistent argument for the role that gas accretion of pristine gas has on regulating star formation in galaxies and the effect that it has in their evolution. The most important aspect of this study is perhaps the evidence that gas flows in galaxies leave imprints in the stellar populations that can be recovered using spectral population fitting techniques and relatively simple models.
The potential for this type of study is evidenced by the fact that we are able to predict the current star formation state of a galaxy given only its past chemical and star formation histories derived by stellar population synthesis.

\subsection{Clues to the origin of the mass-metallicity relation}
The mass-metallicity relation (MZR) is one of the key sources of information regarding how the different evolutionary mechanisms in galaxies interact with one another. Reproducing the shape and values of the MZR has been one of the main ways to test and constrain theoretical models in galaxy evolution \citep[e.g.,][]{Tremonti2004,Kewley2008,Erb2006,Torrey2019,Camps-Farina2021b,Camps-Farina2022}

In a closed box model \citep{Tinsley1974} there is no exchange of material out of the system and the metallicity and stellar mass advance in lockstep as the metal injection is proportional to the stellar mass formed. The predicted mass-metallicity relation is too shallow compared to the measurements which fit with lower mass galaxies having lower effective yields \citep{Tremonti2004}.
The most invoked mechanisms to increase the slope of the MZR are metal-rich outflows and gas accretion as a means to directly reduce the average metallicity of the ISM \citep[e.g.,][]{Larson1974,Tremonti2004,Dalcanton2004,Finlator2008,Zhu2017,Barrera-Ballesteros2018}. There are a number of works, however, which argue that the observed MZR cannot be reproduced solely with accretion \citep{Dalcanton2007} and outflows \citep{Brooks2007,Calura2009}, requiring an additional correlation between the stellar mass and the SFE such that less massive galaxies are less efficient in forming stars. As a result, the amount of metals injected per unit mass of gas would be lower, producing the slope of the MZR.

This hypothesis, however, clashes with observations that low-mass galaxies in the Local Universe tend to have higher SFE than massive ones. On the other hand, our results show that in the past massive galaxies were more efficient at forming stars than low-mass ones, and it is only in recent times that low-mass galaxies became more efficient only because they are more likely to still be forming stars. This is seen in the values of the median SFE and the slope of the SFEH.
Since we also show the effects that gas accretion (and possibly outflows too) has on diluting the ISM over the lifetime of the galaxies our results support a hybrid scenario for the MZR slope to arise from the combination of accretion, outflows as well as a higher SFE for massive galaxies in the past when the bulk of the star formation and therefore metal yields occurred. Other mechanisms such as IMF variation \citep{Koppen2007} with redshift or metallicity as well as mergers \citep{Yates2012} are not discarded by our results due to not being part of the model.

\subsection{Caveats and precision of stellar population synthesis} \label{sec:disc-caveats}
Due to the nature of stellar evolution, the spectrum of a single stellar population changes much more in the time period shortly after being formed compared to when it is of advanced age, due to how the most massive stars die earlier. Because of this, the resolving power for the age of the stellar populations becomes lower as the age increases, which corresponds to lower reliability and resolution of the derived histories at high LBT.

Additionally, stellar population synthesis techniques are well-known to have degeneracies in determining the properties of the populations \citep[e.g.,][]{Walcher2011}, the most important of which is the age-metallicity degeneracy by which an increase in age and an increase in metal abundance have similar effects on the spectra of the templates. As a result, when fitting the populations, the code can obtain a good fit with either (i) the templates of the correct age and metallicity, (ii) templates that are younger but more metallic than the correct ones or (iii) templates that are older but less metallic than the correct ones.

In our testing we have determined that this degeneracy typically produces a positive secondary correlation between SFR and [Z/H] above $\sim$3-5 Gyr in age (see Camps-Fariña et al. in prep.). There are two ways to explain why this is the result: (i) in the LBT range we can resolve, SFHs tend to be decreasing in time and, as such, if the age is underestimated the SFR values are higher than those of the appropriate age. A bias to younger ages is compensated in the degeneracy by using higher abundance templates, thus producing a positive SFR-[Z/H] correlation. Alternatively, we can consider (ii) that the errors in assigning fractions of light to the ages associated with the templates produces an uncertainty in the stellar mass at different LBT. Due to how both the MZR and SFMS have positive correlations with M$_\star$ an overestimation of M$_\star$ means that galaxy is located below both the MZR and SFMS and, conversely, an underestimation of M$_\star$ means the galaxy has higher SFR and [Z/H] than it should for its mass. In either case the result is that the degeneracy produces a positive SFR-[Z/H] correlation.

This is relatively fortunate for the purposes of this work, as the effect of a positive correlation between these parameters will tend to cancel each other to a certain extent. The accretion rate we measure depends mostly on the difference between the expected abundance produced by the yields from the SFR and the measured [Z/H]. If the SFR is increased due to the degeneracy the expected abundance will also increase proportionally. As the [Z/H] will also increase due to the degeneracy, in this case the gap between expected and measured [Z/H] is affected much less than the values of the SFR and [Z/H] themselved, mitigating the effect that the age-metallicity has on the gas accretion rates.

Another way the histories are model dependent is in the choice of templates used to fit the spectra. The range of metallicities that we can measure is directly related to the values that are present in the template grid. The choice of templates is a key aspect of stellar population synthesis as populating the parameter space too much can give rise to significant artifacts at high LBTs if the spectra are too similar to one another (see Discussion section and Appendix B of \citep{Camps-Farina2022}). In the case of templates composed from observed stars, which are more reliable as they do not depend on stellar atmosphere models, finding stars that cover the desired parameter space can also be difficult.

The dependence of the metallicity values on the templates is one of the main reasons that great care should be taken when analyzing our accretion histories as quantitative measurements. The method is capable of finding differences between galaxies even with a relatively narrow range of metallicity values in the templates (see appendices B and C in \cite{Camps-Farina2022}) so the trends are reliable, but a global change to the scaling of the accretion rates can arise if the range of the metallicity in the templates is too narrow. The library employed for this study has a decently wide range of metallicity values with good coverage of low values at Z = 0.0001-0.04.

Overall, we expect a loss of detail in the accretion rates relative to the true ones (i.e., short accretion episodes) due to these effects and the intrinsic capabilities of the method as to how accurately the populations can be recovered. \cite{Ibarra-Medel2019} assessed these effects by taking galaxies from the EAGLE \citep[][]{Schaye2015} simulations and producing mock observations which were then analyzed with Pipe3D, the predecessor to the code we employed. As the "true" SFHs are available from the simulations they can be compared to the ones recovered by the fitting, showing how the histories are similar but the recovered ones lose detail on the individual bursts of star formation \citep[see also][]{Sarmiento2023,Corcho-Caballero2023}.

\subsection{Validity of the physics in the model} \label{sec:disc-valid}
Other than the imprecision intrinsic to the stellar population synthesis method, we are also affected by the significant simplifications of the physical characteristics of the systems we modelled and the processes involved. 
One of the roughest simplifications is the fact that we are effectively considering each galaxy to be a single homogeneous gas cloud into which pristine gas is instantly deposited, on demand. In the case of the averaged histories, we are assuming that this single cloud acts as the average of the galaxies in each bin considered.

In order for this approximation to be valid we expect (i) the averaged histories to be a good proxy for those of an "average galaxy" representative of the bin, and (ii) for the individual histories to be representative of the variety of environments within each galaxy. We are further expecting that the processes that dilute the ISM scale somewhat linearly with other properties such as the stellar mass, SFR and abundance. As an example, consider a galaxy where the chemical enrichment in the central parts is determined solely by the input from stars (closed box) and in the outer parts there is instead significant accretion diluting the content. Our model is valid if the average ChEH combined with the integrated SFH is capable of recovering the entirety of the accretion, which only occurs in the outskirts. In practice we expect this to hold true in principle, provided that we can observe the histories with reasonable precision at each time and cover most of the area of the galaxies. We intend to explore how well this scenario represents the data in future work.

Simulations predict that the bulk of the accretion occurs at the outskirts of the galaxies, so we expect the presence of internal flows to distribute the accreted gas throughout the galaxy. \cite{Genzel2023} report observations on the noncircular motions of CO in galaxies at z$\sim$2, which they use to detect gas which moves inward in the galaxies from its kinematics. Averaging the properties of the galaxies in their table 2 we obtain the following values: log M$_\star$ = 10.95 M$_\odot$, R$_\mathrm{e}$ = 5.8 kpc, f$_\mathrm{gas}$ = 0.52, v$_\mathrm{r}$ = 74 km s$^{-1}$.

If we assume that half the mass of molecular gas is contained within 1 R$_\mathrm{e}$ and apply it to the expression for the total rate of inflow they give: $\dot{M} = \beta \cdot M_\mathrm{gas}(R_\mathrm{e}) \cdot v_r / R_\mathrm{e}$, where $\beta$ is estimated as 0.2 in \cite{Genzel2023}, we obtain an inflow rate of 52 M$_\odot$ yr$^{-1}$ at R$_\mathrm{e}$. This is substantially higher than our corresponding peak accretion rate at this mass range, 5-12 M$_\odot$ yr$^{-1}$ but we have no information about the abundance of the gas as it arrives at 1 R$_\mathrm{e}$ from the outskirts. If the difference in abundance between the gas at the outskirts and at 1 R$_\mathrm{e}$ is much lower than the difference between the latter and value for the accreted gas, we use then it is possible for the dilution effects to be equivalent. Instead of low amounts of very metal poor gas we can match the measured abundance with large amounts of gas with only slightly lower metallicity than the one at 1 R$_\mathrm{e}$. The value of the abundance gradient required for the effect of the inflows from \cite{Genzel2023} to match the dilution caused by the gas accretion we measured is -0.02 dex/kpc for radii larger than 1 R$_\mathrm{e}$.

As such, if the abundance gradient is similar to the inferred value of -0.02 dex kpc$^{-1}$, we can consider the massive inflows measured in \cite{Genzel2023} to represent the inward transport of gas accreted at the outskirts which afterward dilutes the ISM in the entire galaxy.
In this scenario either (i) the gas is quickly mixed once it reaches the galaxy and begins to move inward, (ii) the gas "drags" local clouds inward without mixing with them via destabilizing their orbits or (iii) we are underestimating the abundance of the accreted gas in the first place, lowering our estimations for the accretion rate. The abundance gradient of galaxies at high redshift is poorly constrained, but it is generally expected to be rather flat and our inferred value is very similar to the reported values \citep[e.g.,][]{Swinbank2012,Wuyts2016,Jafariyazani2020}.

Another caveat is that the single cloud which represents the ISM is not divided into the different phases of matter HI, HII and H$_2$, despite their importance in determining the SFH. Regarding the measurement of the accretion histories themselves, this is not critical, and the only question is in the efficiency of the accreted gas to become available for star formation. As we are measuring the accreted gas from its resulting stellar populations, we are intrinsically assuming that 100\% of the accreted gas at least mixes with the gas that will be used to form stars. Any gas which does not reach locations where stars will eventually be formed over the galaxy's lifetime is not observed and therefore our measured accretion histories are lower bounds compared to total the amount of gas that enters the galaxy. The SFE histories, on the other hand, are strongly affected by this, as they are typically measured using the amount (or density) of H$_2$ and are therefore very sensitive to the ratios between phases. In our results this can be seen as we need a conversion factor of 10, consistent with observed HI to H$_2$ ratios, to obtain reasonable values.

The effect of not considering the different phases of matter on the accretion rates therefore depends on which phase the gas mixes in. Consider two edge cases: (i) if the gas efficiently mixes in the HI phase followed by condensation into H$_2$ then our estimates for the gas accretion rate are unaffected by not considering different phases, but (ii) if the gas quickly condenses into H$_2$ and it is only then that the mixing occurs we need to correct by the ratio between M$_\mathrm{HI}$ and M$_\mathrm{H2}$. Following \cite{Calette2018} this ratio should be between 0.1 and 1 so at most it could increase the accretion rate one order of magnitude in the most extreme case of a galaxy in which the accreted gas immediately condenses into H$_2$ and simultaneously has a very small fraction of M$_\mathrm{H2}$ compared to M$_\mathrm{HI}$. This combination is highly unlikely as it requires vastly different condensation timescales for accreted gas and that which is present in the galaxy.

The internal timescales that apply to the journey that the accreted gas travels from the halo until it becomes a star are not considered in the model. Therefore, the accretion histories are expected to be shifted in time by a certain factor. Unless there are significant changes to the physics of the ISM over cosmic time, we can expect this factor to be fairly constant over time, therefore only changing the specific values in the X-axis in the figures, which are already strongly dependent on the choice of isochrones for the stellar population templates. At a minimum, we expect that the delay includes the free-fall time of the cloud from the halo to the galaxy interior, and the timescales for mixing, condensation to H$_2$ and SFR burst.

Another very important simplification is the fact that we do not discriminate between the different effects that can produce the observed "missing metals" in the newer stellar populations. The effect of outflows has been mentioned in the results section to account for some of the observed properties, as their contribution to the gas balance and exchange can significantly alter the results. Depending on the mass-loading factor and the abundance of the ISM at the location where the outflows occur these can either contribute to the dilution, have no effect on the abundance or even produce an over-enrichment compared to the entire galaxy. For example, the centers of galaxies tend to be more metallic than other areas of galaxies, so an AGN outflow with a high mass-loading factor will remove gas that is more metallic than average, thus lowering the average abundance in the galaxy \citep[e.g.,][]{Camps-Farina2021a}. With our method we would conflate the observed drop in expected metallicity to an episode of gas accretion.

Should the metallicity of the gas in an outflow be the same as the average in the galaxy we would not detect anything and if the outflow occurs at the outskirts the gas could be significantly lower in abundance compared to the average in the galaxy, thus raising its average metallicity.
In either of these three cases, we would be overestimating the amount of gas present in the galaxy leading to an underestimation of the SFE.
Most of our results could be recalculated in terms of outflows rather than gas accretion, the assumption of the gas exchange in galaxies being dominated by outflows was explored in \cite{Zhu2017} which presents a leaky-box model where outflows are the primary dilution mechanism managing to reproduce observed resolved properties of MaNGA galaxies. 

In Fig. \ref{fig:ind_mhi} we showed the difference between predicted and measured HI gas mass, and we can use this information to find the mass-loading factor by dividing the difference in gas mass by the integrated SFH. The median mass-loading factor we obtained is 0.96, which is significantly smaller than typical inferred values of 2-3 \citep[e.g.,][]{Bouche2012,Zaragoza-Cardiel2019}. One possible explanation is that the measurements of these studies are biased to concentrated or more violent episodes of star formation while our measurements are averaged over long times and include slower star formation, bringing the average mass-loading factor down. Alternatively, we are underestimating the amount of gas that is accreted to the galaxies, which would increase the mass-loading factor we measured. The most interesting option, however, would be that our measurements are roughly correct and the difference between the values corresponds to the effect of gas that reenters the galaxy after being expelled. Given that we measured the mass-loading factor from the cumulative effects of gas accretion and outflows the time delay to re-accrete the expelled gas is intrinsically included in our calculations. Assuming our measurements are accurate it means that, on average and for the subsample of HI-MaNGA, only about 30-50\% of the gas initially expelled does not fall back into the galaxy after an outflow.

Mergers and interactions between galaxies are also a mechanism that we are not considering, and which can significantly affect both the gas flows within a galaxy and strip it of part of its gas content. Instabilities due to interactions can funnel gas from the outskirts to the center of galaxies and, since the gas at the outskirts tends to have lower abundances, this can mimic an accretion of pristine gas from the halo in our results and probably account for part of the accretion we measured.

The choice of the IMF is a complex topic for the results. We use a Salpeter IMF for consistency with the stellar population synthesis performed on the IFU data. However, our implementation of the SNIa injection (which dominates the Fe yields) is scaled to currently measured rates in the Milky Way and therefore does not explicitly depend on the IMF.
Compared to Salpeter \citep{Salpeter1955}, bottom-light IMFs such as Chabrier \citep{Chabrier2003} or Kroupa \citep{Kroupa1993,Kroupa2001} are typically considered to be a better representation of the true distribution of stars. The difference between these and Salpeter is typically a 60\% factor in stellar mass or SFR owing to the difference in mass to light (M/L) ratios corresponding to each age \citep{Kennicutt2012}. Changing the IMF in our results would reduce the gas accretion rate by a factor between 0 and 60\% as the higher SFR of a Salpeter IMF is balanced by its lower ratio of stars above 1 M$_\odot$. Given main sequence lifetimes, at the present time no stars below 0.9 M$_\odot$ should have left the main sequence and as such the difference in Fe injection between IMFs lies primarily in the amount of stars they predict between 1-8 M$_\odot$ (these are the ones which can produce SNIa). Salpeter IMFs predict higher stellar masses as a result of their higher fraction of stars below 1 M$_\odot$, and as such their fraction of stars between 1-8 M$_\odot$ is lower than for Chabrier or Kroupa IMFs.

Additionally, the validity of a universal IMF shape across cosmic time is highly debated with dissenting results on whether the IMF is top-heavy at high redshift \citep[e.g.,][]{Hayward2013,Eales2023}. In conclusion, the IMF remains a source of uncertainty in the results though its impact in our case is limited due to the fact that our methodology only considers stars with masses above the value at which commonly used IMFs differ.


\section{Conclusions}\label{sec:conclusions}
We have applied a model for the enrichment of the ISM to measurements of SFHs and ChEHs from the MaNGA sample of galaxies. We find that we are able to extract information about the dilution processes that take place in galaxies. We find several trends in gas accretion and evolution of SFE that are in fairly good agreement with expectations from theoretical models as well as good agreement with recent measurements.
Some of the most important results are:
   \begin{enumerate}
      \item We find a good agreement with expected trends for gas accretion in galaxies as a function of their mass, with more massive galaxies having higher peak accretion and a steeper decrease over time.
      \item Controlling for the current stellar mass, galaxies show different accretion histories depending on whether they are forming stars or not. GVG only separate the shape of their average accretion history from SFG in the last Gyr, showing that in general galaxies stop forming stars once they lose access to pristine gas.
      \item The individual accretion histories confirm many of these trends and show a tight correlation between the integrated SFH and the total accreted gas mass. Currently, star-forming galaxies lie above or in the upper envelope of this relation.
      \item The relation between the shape of the accretion history and the current stellar masses and sSFR values indicates that galaxies currently undergoing star formation have persistent gas accretion over cosmic time.
      \item We used the amount of gas predicted to be in the galaxies at each time to calculate their SFE histories. We find that more massive galaxies are currently less efficient, but were more efficient in the past. There is a trend that galaxies with a high median SFE have a steeper decline in SFE over their lifetime.
      \item Galaxies that are currently star-forming tend to have lower SFE at earlier times but persistent SFE over cosmic time.
      \item More compact galaxies are more efficient at forming stars over their lifetimes.
   \end{enumerate}

\begin{acknowledgements}
ACF thanks J. S\'{a}nchez Almeida and C. Dalla Vecchia for useful discussion on the results.
ACF and PSB acknowledge financial support by the Spanish Ministry of Science and Innovation through the research grant PID2019-107427-GB-31. SRF acknowledges financial support from MINECO under grant number AYA2017-90589-REDT, RTI2018-096188-B-I00, and S2018/NMT-429. SFS acknowledges funding from PAPIIT-DGAPA-AG100622 (UNAM) project.

Funding for the Sloan Digital Sky Survey IV has been provided by the Alfred P. Sloan Foundation, the U.S. Department of Energy Office of Science, and the Participating Institutions. SDSS acknowledges support and resources from the Center for High-Performance Computing at the University of Utah. The SDSS web site is \url{www.sdss.org}.

SDSS is managed by the Astrophysical Research Consortium for the Participating Institutions of the SDSS Collaboration including the Brazilian Participation Group, the Carnegie Institution for Science, Carnegie Mellon University, Center for Astrophysics $\vert$ Harvard \& Smithsonian (CfA), the Chilean Participation Group, the French Participation Group, Instituto de Astrof\'{i}sica de Canarias, The Johns Hopkins University, Kavli Institute for the Physics and Mathematics of the Universe (IPMU) / University of Tokyo, the Korean Participation Group, Lawrence Berkeley National Laboratory, Leibniz Institut f\"{u}r Astrophysik Potsdam (AIP), Max-Planck-Institut f\"{u}r Astronomie (MPIA Heidelberg), Max-Planck-Institut f\"{u}r Astrophysik (MPA Garching), Max-Planck-Institut f\"{u}r Extraterrestrische Physik (MPE), National Astronomical Observatories of China, New Mexico State University, New York University, University of Notre Dame, Observat\'{o}rio Nacional / MCTI, The Ohio State University, Pennsylvania State University, Shanghai Astronomical Observatory, United Kingdom Participation Group, Universidad Nacional Aut\'{o}noma de M\'{e}xico, University of Arizona, University of Colorado Boulder, University of Oxford, University of Portsmouth, University of Utah, University of Virginia, University of Washington, University of Wisconsin, Vanderbilt University, and Yale University.

\end{acknowledgements}

%
%
\bibliographystyle{aa} 
\bibliography{export-bibtex.bib} 

\begin{thebibliography}{129}
\expandafter\ifx\csname natexlab\endcsname\relax\def\natexlab#1{#1}\fi

\bibitem[{{Abdurro'uf} {et~al.}(2022){Abdurro'uf}, {Accetta}, {Aerts}, {Silva
  Aguirre}, {Ahumada}, {Ajgaonkar}, {Filiz Ak}, {Alam}, {Allende Prieto},
  {Almeida}, \& et~al.}]{Abdurro'uf2022}
{Abdurro'uf}, {Accetta}, K., {Aerts}, C., {et~al.} 2022, ApJS, 259, 35

\bibitem[{{Baldwin} {et~al.}(1981){Baldwin}, {Phillips}, \&
  {Terlevich}}]{Baldwin1981}
{Baldwin}, J.~A., {Phillips}, M.~M., \& {Terlevich}, R. 1981, PASP, 93, 5

\bibitem[{{Barrera-Ballesteros} {et~al.}(2018){Barrera-Ballesteros}, {Heckman},
  {S{\'a}nchez}, {Zakamska}, {Cleary}, {Zhu}, {Brinkmann}, {Drory}, \& {THE
  MaNGA TEAM}}]{Barrera-Ballesteros2018}
{Barrera-Ballesteros}, J.~K., {Heckman}, T., {S{\'a}nchez}, S.~F., {et~al.}
  2018, ApJ, 852, 74

\bibitem[{{Bouch{\'e}} {et~al.}(2012){Bouch{\'e}}, {Hohensee}, {Vargas},
  {Kacprzak}, {Martin}, {Cooke}, \& {Churchill}}]{Bouche2012}
{Bouch{\'e}}, N., {Hohensee}, W., {Vargas}, R., {et~al.} 2012, MNRAS, 426, 801

\bibitem[{{Bresolin} {et~al.}(2012){Bresolin}, {Kennicutt}, \&
  {Ryan-Weber}}]{Bresolin2012}
{Bresolin}, F., {Kennicutt}, R.~C., \& {Ryan-Weber}, E. 2012, ApJ, 750, 122

\bibitem[{{Brooks} {et~al.}(2007){Brooks}, {Governato}, {Booth}, {Willman},
  {Gardner}, {Wadsley}, {Stinson}, \& {Quinn}}]{Brooks2007}
{Brooks}, A.~M., {Governato}, F., {Booth}, C.~M., {et~al.} 2007, ApJL, 655, L17

\bibitem[{{Bundy} {et~al.}(2015){Bundy}, {Bershady}, {Law}, {Yan}, {Drory},
  {MacDonald}, {Wake}, {Cherinka}, {S{\'a}nchez-Gallego}, {Weijmans}, {Thomas},
  {Tremonti}, {Masters}, {Coccato}, {Diamond-Stanic}, {Arag{\'o}n-Salamanca},
  {Avila-Reese}, {Badenes}, {Falc{\'o}n-Barroso}, {Belfiore}, {Bizyaev},
  {Blanc}, {Bland-Hawthorn}, {Blanton}, {Brownstein}, {Byler}, {Cappellari},
  {Conroy}, {Dutton}, {Emsellem}, {Etherington}, {Frinchaboy}, {Fu}, {Gunn},
  {Harding}, {Johnston}, {Kauffmann}, {Kinemuchi}, {Klaene}, {Knapen},
  {Leauthaud}, {Li}, {Lin}, {Maiolino}, {Malanushenko}, {Malanushenko}, {Mao},
  {Maraston}, {McDermid}, {Merrifield}, {Nichol}, {Oravetz}, {Pan}, {Parejko},
  {Sanchez}, {Schlegel}, {Simmons}, {Steele}, {Steinmetz}, {Thanjavur},
  {Thompson}, {Tinker}, {van den Bosch}, {Westfall}, {Wilkinson}, {Wright},
  {Xiao}, \& {Zhang}}]{Bundy2015}
{Bundy}, K., {Bershady}, M.~A., {Law}, D.~R., {et~al.} 2015, ApJ, 798, 7

\bibitem[{{Caimmi}(2008)}]{Caimmi2008}
{Caimmi}, R. 2008, NewA, 13, 314

\bibitem[{{Calette} {et~al.}(2018){Calette}, {Avila-Reese},
  {Rodr{\'\i}guez-Puebla}, {Hern{\'a}ndez-Toledo}, \&
  {Papastergis}}]{Calette2018}
{Calette}, A.~R., {Avila-Reese}, V., {Rodr{\'\i}guez-Puebla}, A.,
  {Hern{\'a}ndez-Toledo}, H., \& {Papastergis}, E. 2018, RMxAA, 54, 443

\bibitem[{{Calura} {et~al.}(2009){Calura}, {Pipino}, {Chiappini}, {Matteucci},
  \& {Maiolino}}]{Calura2009}
{Calura}, F., {Pipino}, A., {Chiappini}, C., {Matteucci}, F., \& {Maiolino}, R.
  2009, A\&A, 504, 373

\bibitem[{{Camps-Fari{\~n}a} {et~al.}(2021{\natexlab{a}}){Camps-Fari{\~n}a},
  {S{\'a}nchez}, {Carigi}, {Lacerda}, {Garc{\'\i}a-Benito}, {Mast}, {Galbany},
  \& {Barrera-Ballesteros}}]{Camps-Farina2021a}
{Camps-Fari{\~n}a}, A., {S{\'a}nchez}, S.~F., {Carigi}, L., {et~al.}
  2021{\natexlab{a}}, ApJL, 922, L20

\bibitem[{{Camps-Fari{\~n}a} {et~al.}(2021{\natexlab{b}}){Camps-Fari{\~n}a},
  {Sanchez}, {Lacerda}, {Carigi}, {Garc{\'\i}a-Benito}, {Mast}, \&
  {Galbany}}]{Camps-Farina2021b}
{Camps-Fari{\~n}a}, A., {Sanchez}, S.~F., {Lacerda}, E.~A.~D., {et~al.}
  2021{\natexlab{b}}, MNRAS, 504, 3478

\bibitem[{{Camps-Fari{\~n}a} {et~al.}(2022){Camps-Fari{\~n}a}, {S{\'a}nchez},
  {Mej{\'\i}a-Narv{\'a}ez}, {Lacerda}, {Carigi}, {Bruzual}, {Alvarez-Hurtado},
  {Drory}, {Lane}, {Boardman}, \& {Blanc}}]{Camps-Farina2022}
{Camps-Fari{\~n}a}, A., {S{\'a}nchez}, S.~F., {Mej{\'\i}a-Narv{\'a}ez}, A.,
  {et~al.} 2022, ApJ, 933, 44

\bibitem[{{Ceverino} {et~al.}(2016){Ceverino}, {S{\'a}nchez Almeida},
  {Mu{\~n}oz Tu{\~n}{\'o}n}, {Dekel}, {Elmegreen}, {Elmegreen}, \&
  {Primack}}]{Ceverino2016}
{Ceverino}, D., {S{\'a}nchez Almeida}, J., {Mu{\~n}oz Tu{\~n}{\'o}n}, C.,
  {et~al.} 2016, MNRAS, 457, 2605

\bibitem[{{Chabrier}(2003)}]{Chabrier2003}
{Chabrier}, G. 2003, PASP, 115, 763

\bibitem[{{Chiappini}(2009)}]{Chiappini2009}
{Chiappini}, C. 2009, in The Galaxy Disk in Cosmological Context, ed.
  J.~{Andersen}, {Nordstr{\"o}ara}, B.~{m}, \& J.~{Bland-Hawthorn}, Vol. 254,
  191--196

\bibitem[{{Chiappini} {et~al.}(1997){Chiappini}, {Matteucci}, \&
  {Gratton}}]{Chiappini1997}
{Chiappini}, C., {Matteucci}, F., \& {Gratton}, R. 1997, ApJ, 477, 765

\bibitem[{{Chisholm} {et~al.}(2017){Chisholm}, {Orlitov{\'a}}, {Schaerer},
  {Verhamme}, {Worseck}, {Izotov}, {Thuan}, \& {Guseva}}]{Chisholm2017}
{Chisholm}, J., {Orlitov{\'a}}, I., {Schaerer}, D., {et~al.} 2017, A\&A, 605,
  A67

\bibitem[{{Chowdhury} {et~al.}(2022){Chowdhury}, {Kanekar}, \&
  {Chengalur}}]{Chowdhury2022}
{Chowdhury}, A., {Kanekar}, N., \& {Chengalur}, J.~N. 2022, ApJL, 941, L6

\bibitem[{{Cid Fernandes} {et~al.}(2014){Cid Fernandes}, {Gonz{\'a}lez
  Delgado}, {Garc{\'\i}a Benito}, {P{\'e}rez}, {de Amorim}, {S{\'a}nchez},
  {Husemann}, {Falc{\'o}n Barroso}, {L{\'o}pez-Fern{\'a}ndez},
  {S{\'a}nchez-Bl{\'a}zquez}, {Vale Asari}, {Vazdekis}, {Walcher}, \&
  {Mast}}]{CidFernandes2014}
{Cid Fernandes}, R., {Gonz{\'a}lez Delgado}, R.~M., {Garc{\'\i}a Benito}, R.,
  {et~al.} 2014, A\&A, 561, A130

\bibitem[{{Cid Fernandes} {et~al.}(2011){Cid Fernandes}, {Stasi{\'n}ska},
  {Mateus}, \& {Vale Asari}}]{CidFernandes2011}
{Cid Fernandes}, R., {Stasi{\'n}ska}, G., {Mateus}, A., \& {Vale Asari}, N.
  2011, MNRAS, 413, 1687

\bibitem[{{Cimatti} {et~al.}(2019){Cimatti}, {Fraternali}, \&
  {Nipoti}}]{Cimatti2019}
{Cimatti}, A., {Fraternali}, F., \& {Nipoti}, C. 2019, {Introduction to Galaxy
  Formation and Evolution: From Primordial Gas to Present-Day Galaxies}

\bibitem[{{Colombo} {et~al.}(2018){Colombo}, {Kalinova}, {Utomo}, {Rosolowsky},
  {Bolatto}, {Levy}, {Wong}, {Sanchez}, {Leroy}, {Ostriker}, {Blitz}, {Vogel},
  {Mast}, {Garc{\'\i}a-Benito}, {Husemann}, {Dannerbauer}, {Ellmeier}, \&
  {Cao}}]{Colombo2018}
{Colombo}, D., {Kalinova}, V., {Utomo}, D., {et~al.} 2018, MNRAS, 475, 1791

\bibitem[{{Colombo} {et~al.}(2020){Colombo}, {Sanchez}, {Bolatto}, {Kalinova},
  {Wei{\ss}}, {Wong}, {Rosolowsky}, {Vogel}, {Barrera-Ballesteros},
  {Dannerbauer}, {Cao}, {Levy}, {Utomo}, \& {Blitz}}]{Colombo2020}
{Colombo}, D., {Sanchez}, S.~F., {Bolatto}, A.~D., {et~al.} 2020, A\&A, 644,
  A97

\bibitem[{{Concas} {et~al.}(2022){Concas}, {Maiolino}, {Curti},
  {Hayden-Pawson}, {Cirasuolo}, {Jones}, {Mercurio}, {Belfiore}, {Cresci},
  {Cullen}, {Mannucci}, {Marconi}, {Cappellari}, {Cicone}, {Peng}, \&
  {Troncoso}}]{Concas2022}
{Concas}, A., {Maiolino}, R., {Curti}, M., {et~al.} 2022, MNRAS, 513, 2535

\bibitem[{{Corcho-Caballero} {et~al.}(2023){Corcho-Caballero}, {Ascasibar},
  {S{\'a}nchez}, \& {L{\'o}pez-S{\'a}nchez}}]{Corcho-Caballero2023}
{Corcho-Caballero}, P., {Ascasibar}, Y., {S{\'a}nchez}, S.~F., \&
  {L{\'o}pez-S{\'a}nchez}, {\'A}.~R. 2023, MNRAS, 520, 193

\bibitem[{{Cresci} {et~al.}(2010){Cresci}, {Mannucci}, {Maiolino}, {Marconi},
  {Gnerucci}, \& {Magrini}}]{Cresci2010}
{Cresci}, G., {Mannucci}, F., {Maiolino}, R., {et~al.} 2010, Natur, 467, 811

\bibitem[{{Daddi} {et~al.}(2010){Daddi}, {Bournaud}, {Walter}, {Dannerbauer},
  {Carilli}, {Dickinson}, {Elbaz}, {Morrison}, {Riechers}, {Onodera}, {Salmi},
  {Krips}, \& {Stern}}]{Daddi2010}
{Daddi}, E., {Bournaud}, F., {Walter}, F., {et~al.} 2010, ApJ, 713, 686

\bibitem[{{Dalcanton}(2007)}]{Dalcanton2007}
{Dalcanton}, J.~J. 2007, ApJ, 658, 941

\bibitem[{{Dalcanton} {et~al.}(2004){Dalcanton}, {Yoachim}, \&
  {Bernstein}}]{Dalcanton2004}
{Dalcanton}, J.~J., {Yoachim}, P., \& {Bernstein}, R.~A. 2004, ApJ, 608, 189

\bibitem[{{Davis} \& {Bureau}(2016)}]{Davis2016}
{Davis}, T.~A. \& {Bureau}, M. 2016, MNRAS, 457, 272

\bibitem[{{Dekel} \& {Birnboim}(2006)}]{Dekel2006}
{Dekel}, A. \& {Birnboim}, Y. 2006, MNRAS, 368, 2

\bibitem[{{Dekel} {et~al.}(2009){Dekel}, {Birnboim}, {Engel}, {Freundlich},
  {Goerdt}, {Mumcuoglu}, {Neistein}, {Pichon}, {Teyssier}, \&
  {Zinger}}]{Dekel2009}
{Dekel}, A., {Birnboim}, Y., {Engel}, G., {et~al.} 2009, Natur, 457, 451

\bibitem[{{Drory} {et~al.}(2015){Drory}, {MacDonald}, {Bershady}, {Bundy},
  {Gunn}, {Law}, {Smith}, {Stoll}, {Tremonti}, {Wake}, {Yan}, {Weijmans},
  {Byler}, {Cherinka}, {Cope}, {Eigenbrot}, {Harding}, {Holder}, {Huehnerhoff},
  {Jaehnig}, {Jansen}, {Klaene}, {Paat}, {Percival}, \& {Sayres}}]{Drory2015}
{Drory}, N., {MacDonald}, N., {Bershady}, M.~A., {et~al.} 2015, AJ, 149, 77

\bibitem[{{Eales} {et~al.}(2023){Eales}, {Gomez}, {Dunne}, {Dye}, \&
  {Smith}}]{Eales2023}
{Eales}, S., {Gomez}, H., {Dunne}, L., {Dye}, S., \& {Smith}, M. W.~L. 2023,
  arXiv, arXiv:2303.07376

\bibitem[{{Erb} {et~al.}(2006){Erb}, {Shapley}, {Pettini}, {Steidel}, {Reddy},
  \& {Adelberger}}]{Erb2006}
{Erb}, D.~K., {Shapley}, A.~E., {Pettini}, M., {et~al.} 2006, ApJ, 644, 813

\bibitem[{{Fenner} \& {Gibson}(2003)}]{Fenner2003}
{Fenner}, Y. \& {Gibson}, B.~K. 2003, PASA, 20, 189

\bibitem[{{Finlator} \& {Dav{\'e}}(2008)}]{Finlator2008}
{Finlator}, K. \& {Dav{\'e}}, R. 2008, MNRAS, 385, 2181

\bibitem[{{Fraternali} \& {Binney}(2008)}]{Fraternali2008}
{Fraternali}, F. \& {Binney}, J.~J. 2008, MNRAS, 386, 935

\bibitem[{{Fraternali} \& {Tomassetti}(2012)}]{Fraternali2012}
{Fraternali}, F. \& {Tomassetti}, M. 2012, MNRAS, 426, 2166

\bibitem[{{Genzel} {et~al.}(2014){Genzel}, {F{\"o}rster Schreiber}, {Rosario},
  {Lang}, {Lutz}, {Wisnioski}, {Wuyts}, {Wuyts}, {Bandara}, {Bender}, {Berta},
  {Kurk}, {Mendel}, {Tacconi}, {Wilman}, {Beifiori}, {Brammer}, {Burkert},
  {Buschkamp}, {Chan}, {Carollo}, {Davies}, {Eisenhauer}, {Fabricius},
  {Fossati}, {Kriek}, {Kulkarni}, {Lilly}, {Mancini}, {Momcheva}, {Naab},
  {Nelson}, {Renzini}, {Saglia}, {Sharples}, {Sternberg}, {Tacchella}, \& {van
  Dokkum}}]{Genzel2014}
{Genzel}, R., {F{\"o}rster Schreiber}, N.~M., {Rosario}, D., {et~al.} 2014,
  ApJ, 796, 7

\bibitem[{{Genzel} {et~al.}(2023){Genzel}, {Jolly}, {Liu}, {Price},
  {F{\"o}rster Schreiber}, {Tacconi}, {Herrera-Camus}, {Barfety}, {Burkert},
  {Cao}, {Davies}, {Dekel}, {Lee}, {Lee}, {Lutz}, {Naab}, {Neri}, {Nestor
  Shachar}, {Pastras}, {Pulsoni}, {Renzini}, {Schuster}, {Shimizu}, {Stanley},
  {Sternberg}, \& {{\"U}bler}}]{Genzel2023}
{Genzel}, R., {Jolly}, J.~B., {Liu}, D., {et~al.} 2023, arXiv, arXiv:2305.02959

\bibitem[{{Genzel} {et~al.}(2010){Genzel}, {Tacconi}, {Gracia-Carpio},
  {Sternberg}, {Cooper}, {Shapiro}, {Bolatto}, {Bouch{\'e}}, {Bournaud},
  {Burkert}, {Combes}, {Comerford}, {Cox}, {Davis}, {F{\"o}rster Schreiber},
  {Garcia-Burillo}, {Lutz}, {Naab}, {Neri}, {Omont}, {Shapley}, \&
  {Weiner}}]{Genzel2010}
{Genzel}, R., {Tacconi}, L.~J., {Gracia-Carpio}, J., {et~al.} 2010, MNRAS, 407,
  2091

\bibitem[{{Giavalisco} {et~al.}(2011){Giavalisco}, {Vanzella}, {Salimbeni},
  {Tripp}, {Dickinson}, {Cassata}, {Renzini}, {Guo}, {Ferguson}, {Nonino},
  {Cimatti}, {Kurk}, {Mignoli}, \& {Tang}}]{Giavalisco2011}
{Giavalisco}, M., {Vanzella}, E., {Salimbeni}, S., {et~al.} 2011, ApJ, 743, 95

\bibitem[{{Gonz{\'a}lez Delgado} {et~al.}(2014){Gonz{\'a}lez Delgado}, {Cid
  Fernandes}, {Garc{\'\i}a-Benito}, {P{\'e}rez}, {de Amorim},
  {Cortijo-Ferrero}, {Lacerda}, {L{\'o}pez Fern{\'a}ndez}, {S{\'a}nchez}, {Vale
  Asari}, {Alves}, {Bland-Hawthorn}, {Galbany}, {Gallazzi}, {Husemann},
  {Bekeraite}, {Jungwiert}, {L{\'o}pez-S{\'a}nchez}, {de Lorenzo-C{\'a}ceres},
  {Marino}, {Mast}, {Moll{\'a}}, {del Olmo}, {S{\'a}nchez-Bl{\'a}zquez}, {van
  de Ven}, {V{\'\i}lchez}, {Walcher}, {Wisotzki}, {Ziegler}, \& {CALIFA
  Collaboration}}]{GonzalezDelgado2014}
{Gonz{\'a}lez Delgado}, R.~M., {Cid Fernandes}, R., {Garc{\'\i}a-Benito}, R.,
  {et~al.} 2014, ApJL, 791, L16

\bibitem[{{Greggio}(2010)}]{Greggio2010}
{Greggio}, L. 2010, MNRAS, 406, 22

\bibitem[{{Gunn} {et~al.}(2006){Gunn}, {Siegmund}, {Mannery}, {Owen}, {Hull},
  {Leger}, {Carey}, {Knapp}, {York}, {Boroski}, {Kent}, {Lupton}, {Rockosi},
  {Evans}, {Waddell}, {Anderson}, {Annis}, {Barentine}, {Bartoszek}, {Bastian},
  {Bracker}, {Brewington}, {Briegel}, {Brinkmann}, {Brown}, {Carr},
  {Czarapata}, {Drennan}, {Dombeck}, {Federwitz}, {Gillespie}, {Gonzales},
  {Hansen}, {Harvanek}, {Hayes}, {Jordan}, {Kinney}, {Klaene}, {Kleinman},
  {Kron}, {Kresinski}, {Lee}, {Limmongkol}, {Lindenmeyer}, {Long}, {Loomis},
  {McGehee}, {Mantsch}, {Neilsen}, {Neswold}, {Newman}, {Nitta}, {Peoples},
  {Pier}, {Prieto}, {Prosapio}, {Rivetta}, {Schneider}, {Snedden}, \&
  {Wang}}]{Gunn2006}
{Gunn}, J.~E., {Siegmund}, W.~A., {Mannery}, E.~J., {et~al.} 2006, AJ, 131,
  2332

\bibitem[{{Hayward} {et~al.}(2013){Hayward}, {Narayanan}, {Kere{\v{s}}},
  {Jonsson}, {Hopkins}, {Cox}, \& {Hernquist}}]{Hayward2013}
{Hayward}, C.~C., {Narayanan}, D., {Kere{\v{s}}}, D., {et~al.} 2013, MNRAS,
  428, 2529

\bibitem[{{Heintz} {et~al.}(2022){Heintz}, {Oesch}, {Aravena}, {Bouwens},
  {Dayal}, {Ferrara}, {Fudamoto}, {Graziani}, {Inami}, {Sommovigo}, {Smit},
  {Stefanon}, {Topping}, {Pallottini}, \& {van der Werf}}]{Heintz2022}
{Heintz}, K.~E., {Oesch}, P.~A., {Aravena}, M., {et~al.} 2022, ApJL, 934, L27

\bibitem[{{Hernanz}(2005)}]{Hernanz2005}
{Hernanz}, M. 2005, in Astronomical Society of the Pacific Conference Series,
  Vol. 330, The Astrophysics of Cataclysmic Variables and Related Objects, ed.
  J.~M. {Hameury} \& J.~P. {Lasota}, 265

\bibitem[{{Hillman} {et~al.}(2015){Hillman}, {Prialnik}, {Kovetz}, \&
  {Shara}}]{Hillman2015}
{Hillman}, Y., {Prialnik}, D., {Kovetz}, A., \& {Shara}, M.~M. 2015, MNRAS,
  446, 1924

\bibitem[{{Ibarra-Medel} {et~al.}(2019){Ibarra-Medel}, {Avila-Reese},
  {S{\'a}nchez}, {Gonz{\'a}lez-Samaniego}, \&
  {Rodr{\'\i}guez-Puebla}}]{Ibarra-Medel2019}
{Ibarra-Medel}, H.~J., {Avila-Reese}, V., {S{\'a}nchez}, S.~F.,
  {Gonz{\'a}lez-Samaniego}, A., \& {Rodr{\'\i}guez-Puebla}, A. 2019, MNRAS,
  483, 4525

\bibitem[{{Izzo} {et~al.}(2015){Izzo}, {Della Valle}, {Mason}, {Matteucci},
  {Romano}, {Pasquini}, {Vanzi}, {Jordan}, {Fernandez}, {Bluhm}, {Brahm},
  {Espinoza}, \& {Williams}}]{Izzo2015}
{Izzo}, L., {Della Valle}, M., {Mason}, E., {et~al.} 2015, ApJL, 808, L14

\bibitem[{{Jafariyazani} {et~al.}(2020){Jafariyazani}, {Newman}, {Mobasher},
  {Belli}, {Ellis}, \& {Patel}}]{Jafariyazani2020}
{Jafariyazani}, M., {Newman}, A.~B., {Mobasher}, B., {et~al.} 2020, ApJL, 897,
  L42

\bibitem[{{Kado-Fong} {et~al.}(2020){Kado-Fong}, {Kim}, {Ostriker}, \&
  {Kim}}]{Kado-Fong2020}
{Kado-Fong}, E., {Kim}, J.-G., {Ostriker}, E.~C., \& {Kim}, C.-G. 2020, ApJ,
  897, 143

\bibitem[{{Kennicutt}(1983)}]{Kennicutt1983}
{Kennicutt}, R.~C., J. 1983, ApJ, 272, 54

\bibitem[{{Kennicutt}(1998)}]{Kennicutt1998}
{Kennicutt}, Robert~C., J. 1998, ApJ, 498, 541

\bibitem[{{Kennicutt} \& {Evans}(2012)}]{Kennicutt2012}
{Kennicutt}, R.~C. \& {Evans}, N.~J. 2012, ARA\&A, 50, 531

\bibitem[{{Kere{\v{s}}} {et~al.}(2005){Kere{\v{s}}}, {Katz}, {Weinberg}, \&
  {Dav{\'e}}}]{Keres2005}
{Kere{\v{s}}}, D., {Katz}, N., {Weinberg}, D.~H., \& {Dav{\'e}}, R. 2005,
  MNRAS, 363, 2

\bibitem[{{Kewley} \& {Ellison}(2008)}]{Kewley2008}
{Kewley}, L.~J. \& {Ellison}, S.~L. 2008, ApJ, 681, 1183

\bibitem[{{Kobayashi} {et~al.}(2011){Kobayashi}, {Karakas}, \&
  {Umeda}}]{Kobayashi2011}
{Kobayashi}, C., {Karakas}, A.~I., \& {Umeda}, H. 2011, MNRAS, 414, 3231

\bibitem[{{K{\"o}ppen} {et~al.}(2007){K{\"o}ppen}, {Weidner}, \&
  {Kroupa}}]{Koppen2007}
{K{\"o}ppen}, J., {Weidner}, C., \& {Kroupa}, P. 2007, MNRAS, 375, 673

\bibitem[{{Kroupa}(2001)}]{Kroupa2001}
{Kroupa}, P. 2001, MNRAS, 322, 231

\bibitem[{{Kroupa} {et~al.}(1993){Kroupa}, {Tout}, \& {Gilmore}}]{Kroupa1993}
{Kroupa}, P., {Tout}, C.~A., \& {Gilmore}, G. 1993, MNRAS, 262, 545

\bibitem[{{Lacerda} {et~al.}(2020){Lacerda}, {S{\'a}nchez}, {Cid Fernandes},
  {L{\'o}pez-Cob{\'a}}, {Espinosa-Ponce}, \& {Galbany}}]{Lacerda2020}
{Lacerda}, E. A.~D., {S{\'a}nchez}, S.~F., {Cid Fernandes}, R., {et~al.} 2020,
  MNRAS, 492, 3073

\bibitem[{{Lacerda} {et~al.}(2022){Lacerda}, {S{\'a}nchez},
  {Mej{\'\i}a-Narv{\'a}ez}, {Camps-Fari{\~n}a}, {Espinosa-Ponce},
  {Barrera-Ballesteros}, {Ibarra-Medel}, \& {Lugo-Aranda}}]{Lacerda2022}
{Lacerda}, E. A.~D., {S{\'a}nchez}, S.~F., {Mej{\'\i}a-Narv{\'a}ez}, A.,
  {et~al.} 2022, NewA, 97, 101895

\bibitem[{{Lah} {et~al.}(2007){Lah}, {Chengalur}, {Briggs}, {Colless}, {de
  Propris}, {Pracy}, {de Blok}, {Fujita}, {Ajiki}, {Shioya}, {Nagao},
  {Murayama}, {Taniguchi}, {Yagi}, \& {Okamura}}]{Lah2007}
{Lah}, P., {Chengalur}, J.~N., {Briggs}, F.~H., {et~al.} 2007, MNRAS, 376, 1357

\bibitem[{{Larson}(1972)}]{Larson1972}
{Larson}, R.~B. 1972, NPhS, 236, 7

\bibitem[{{Larson}(1974)}]{Larson1974}
{Larson}, R.~B. 1974, MNRAS, 169, 229

\bibitem[{{Law} {et~al.}(2016){Law}, {Cherinka}, {Yan}, {Andrews}, {Bershady},
  {Bizyaev}, {Blanc}, {Blanton}, {Bolton}, {Brownstein}, {Bundy}, {Chen},
  {Drory}, {D'Souza}, {Fu}, {Jones}, {Kauffmann}, {MacDonald}, {Masters},
  {Newman}, {Parejko}, {S{\'a}nchez-Gallego}, {S{\'a}nchez}, {Schlegel},
  {Thomas}, {Wake}, {Weijmans}, {Westfall}, \& {Zhang}}]{Law2016}
{Law}, D.~R., {Cherinka}, B., {Yan}, R., {et~al.} 2016, AJ, 152, 83

\bibitem[{{Leroy} {et~al.}(2008){Leroy}, {Walter}, {Brinks}, {Bigiel}, {de
  Blok}, {Madore}, \& {Thornley}}]{Leroy2008}
{Leroy}, A.~K., {Walter}, F., {Brinks}, E., {et~al.} 2008, AJ, 136, 2782

\bibitem[{{Lilly} {et~al.}(2013){Lilly}, {Carollo}, {Pipino}, {Renzini}, \&
  {Peng}}]{Lilly2013}
{Lilly}, S.~J., {Carollo}, C.~M., {Pipino}, A., {Renzini}, A., \& {Peng}, Y.
  2013, ApJ, 772, 119

\bibitem[{{L{\'o}pez-Cob{\'a}} {et~al.}(2020){L{\'o}pez-Cob{\'a}},
  {S{\'a}nchez}, {Anderson}, {Cruz-Gonz{\'a}lez}, {Galbany}, {Ruiz-Lara},
  {Barrera-Ballesteros}, {Prieto}, \& {Kuncarayakti}}]{Lopez-Coba2020}
{L{\'o}pez-Cob{\'a}}, C., {S{\'a}nchez}, S.~F., {Anderson}, J.~P., {et~al.}
  2020, AJ, 159, 167

\bibitem[{{L{\'o}pez-Cob{\'a}} {et~al.}(2019){L{\'o}pez-Cob{\'a}},
  {S{\'a}nchez}, {Bland-Hawthorn}, {Moiseev}, {Cruz-Gonz{\'a}lez},
  {Garc{\'\i}a-Benito}, {Barrera-Ballesteros}, \& {Galbany}}]{Lopez-Coba2019}
{L{\'o}pez-Cob{\'a}}, C., {S{\'a}nchez}, S.~F., {Bland-Hawthorn}, J., {et~al.}
  2019, MNRAS, 482, 4032

\bibitem[{{Martin} {et~al.}(2012){Martin}, {Shapley}, {Coil}, {Kornei},
  {Bundy}, {Weiner}, {Noeske}, \& {Schiminovich}}]{Martin2012}
{Martin}, C.~L., {Shapley}, A.~E., {Coil}, A.~L., {et~al.} 2012, ApJ, 760, 127

\bibitem[{{Masters} {et~al.}(2019){Masters}, {Stark}, {Pace}, {Phipps},
  {Rujopakarn}, {Samanso}, {Harrington}, {S{\'a}nchez-Gallego}, {Avila-Reese},
  {Bershady}, {Cherinka}, {Fielder}, {Finnegan}, {Riffel}, {Rowlands},
  {Shamsi}, {Newnham}, {Weijmans}, \& {Witherspoon}}]{Masters2019}
{Masters}, K.~L., {Stark}, D.~V., {Pace}, Z.~J., {et~al.} 2019, MNRAS, 488,
  3396

\bibitem[{{Minchev} {et~al.}(2018){Minchev}, {Anders}, {Recio-Blanco},
  {Chiappini}, {de Laverny}, {Queiroz}, {Steinmetz}, {Adibekyan}, {Carrillo},
  {Cescutti}, {Guiglion}, {Hayden}, {de Jong}, {Kordopatis}, {Majewski},
  {Martig}, \& {Santiago}}]{Minchev2018}
{Minchev}, I., {Anders}, F., {Recio-Blanco}, A., {et~al.} 2018, MNRAS, 481,
  1645

\bibitem[{{Mitchell} {et~al.}(2020){Mitchell}, {Schaye}, {Bower}, \&
  {Crain}}]{Mitchell2020}
{Mitchell}, P.~D., {Schaye}, J., {Bower}, R.~G., \& {Crain}, R.~A. 2020, MNRAS,
  494, 3971

\bibitem[{{Moll{\'a}} {et~al.}(2016){Moll{\'a}}, {D{\'\i}az}, {Gibson},
  {Cavichia}, \& {L{\'o}pez-S{\'a}nchez}}]{Molla2016}
{Moll{\'a}}, M., {D{\'\i}az}, {\'A}.~I., {Gibson}, B.~K., {Cavichia}, O., \&
  {L{\'o}pez-S{\'a}nchez}, {\'A}.-R. 2016, MNRAS, 462, 1329

\bibitem[{{Muzzin} {et~al.}(2009){Muzzin}, {Marchesini}, {van Dokkum},
  {Labb{\'e}}, {Kriek}, \& {Franx}}]{Muzzin2009}
{Muzzin}, A., {Marchesini}, D., {van Dokkum}, P.~G., {et~al.} 2009, ApJ, 701,
  1839

\bibitem[{{Nordstr{\"o}m} {et~al.}(2004){Nordstr{\"o}m}, {Mayor}, {Andersen},
  {Holmberg}, {Pont}, {J{\o}rgensen}, {Olsen}, {Udry}, \&
  {Mowlavi}}]{Nordstrom2004}
{Nordstr{\"o}m}, B., {Mayor}, M., {Andersen}, J., {et~al.} 2004, A\&A, 418, 989

\bibitem[{{Parkash} {et~al.}(2018){Parkash}, {Brown}, {Jarrett}, \&
  {Bonne}}]{Parkash2018}
{Parkash}, V., {Brown}, M. J.~I., {Jarrett}, T.~H., \& {Bonne}, N.~J. 2018,
  ApJ, 864, 40

\bibitem[{{Poudel} {et~al.}(2020){Poudel}, {Kulkarni}, {Cashman}, {Frye},
  {P{\'e}roux}, {Rahmani}, \& {Quiret}}]{Poudel2020}
{Poudel}, S., {Kulkarni}, V.~P., {Cashman}, F.~H., {et~al.} 2020, MNRAS, 491,
  1008

\bibitem[{{Prochaska} {et~al.}(2005){Prochaska}, {Herbert-Fort}, \&
  {Wolfe}}]{Prochaska2005}
{Prochaska}, J.~X., {Herbert-Fort}, S., \& {Wolfe}, A.~M. 2005, ApJ, 635, 123

\bibitem[{{Putman} {et~al.}(2012){Putman}, {Peek}, \& {Joung}}]{Putman2012}
{Putman}, M.~E., {Peek}, J.~E.~G., \& {Joung}, M.~R. 2012, ARA\&A, 50, 491

\bibitem[{{Rao} {et~al.}(2006){Rao}, {Turnshek}, \& {Nestor}}]{Rao2006}
{Rao}, S.~M., {Turnshek}, D.~A., \& {Nestor}, D.~B. 2006, ApJ, 636, 610

\bibitem[{{Roca-F{\`a}brega} {et~al.}(2021){Roca-F{\`a}brega}, {Llorente de
  Andr{\'e}s}, {Chavero}, {Cifuentes}, \& {de la Reza}}]{Roca-Fabrega2021}
{Roca-F{\`a}brega}, S., {Llorente de Andr{\'e}s}, F., {Chavero}, C.,
  {Cifuentes}, C., \& {de la Reza}, R. 2021, A\&A, 656, A64

\bibitem[{{Rodr{\'\i}guez-Puebla} {et~al.}(2016){Rodr{\'\i}guez-Puebla},
  {Primack}, {Behroozi}, \& {Faber}}]{Rodriguez-Puebla2016}
{Rodr{\'\i}guez-Puebla}, A., {Primack}, J.~R., {Behroozi}, P., \& {Faber},
  S.~M. 2016, MNRAS, 455, 2592

\bibitem[{{Rubin} {et~al.}(2012){Rubin}, {Prochaska}, {Koo}, \&
  {Phillips}}]{Rubin2012}
{Rubin}, K. H.~R., {Prochaska}, J.~X., {Koo}, D.~C., \& {Phillips}, A.~C. 2012,
  ApJL, 747, L26

\bibitem[{{Salpeter}(1955)}]{Salpeter1955}
{Salpeter}, E.~E. 1955, ApJ, 121, 161

\bibitem[{{S{\'a}nchez}(2006)}]{Sanchez2006}
{S{\'a}nchez}, S.~F. 2006, AN, 327, 850

\bibitem[{{S{\'a}nchez}(2020)}]{Sanchez2020}
{S{\'a}nchez}, S.~F. 2020, ARA\&A, 58, 99

\bibitem[{{S{\'a}nchez} {et~al.}(2021{\natexlab{a}}){S{\'a}nchez},
  {Barrera-Ballesteros}, {Colombo}, {Wong}, {Bolatto}, {Rosolowsky}, {Vogel},
  {Levy}, {Kalinova}, {Alvarez-Hurtado}, {Luo}, \& {Cao}}]{Sanchez2021b}
{S{\'a}nchez}, S.~F., {Barrera-Ballesteros}, J.~K., {Colombo}, D., {et~al.}
  2021{\natexlab{a}}, MNRAS, 503, 1615

\bibitem[{{S{\'a}nchez} {et~al.}(2022){S{\'a}nchez}, {Barrera-Ballesteros},
  {Lacerda}, {Mej{\'\i}a-Narvaez}, {Camps-Fari{\~n}a}, {Bruzual},
  {Espinosa-Ponce}, {Rodr{\'\i}guez-Puebla}, {Calette}, {Ibarra-Medel},
  {Avila-Reese}, {Hernandez-Toledo}, {Bershady}, {Cano-Diaz}, \&
  {Munguia-Cordova}}]{Sanchez2022}
{S{\'a}nchez}, S.~F., {Barrera-Ballesteros}, J.~K., {Lacerda}, E., {et~al.}
  2022, ApJS, 262, 36

\bibitem[{{S{\'a}nchez} {et~al.}(2021{\natexlab{b}}){S{\'a}nchez},
  {Espinosa-Ponce}, {Carigi}, {Morisset}, {Barrera-Ballesteros}, {Walcher},
  {Garc{\'\i}a-Benito}, {Camps-Fari{\~n}a}, \& {Galbany}}]{Sanchez2021a}
{S{\'a}nchez}, S.~F., {Espinosa-Ponce}, C., {Carigi}, L., {et~al.}
  2021{\natexlab{b}}, A\&A, 652, L10

\bibitem[{{S{\'a}nchez} {et~al.}(2016{\natexlab{a}}){S{\'a}nchez}, {P{\'e}rez},
  {S{\'a}nchez-Bl{\'a}zquez}, {Garc{\'\i}a-Benito}, {Ibarra-Mede},
  {Gonz{\'a}lez}, {Rosales-Ortega}, {S{\'a}nchez-Menguiano}, {Ascasibar},
  {Bitsakis}, {Law}, {Cano-D{\'\i}az}, {L{\'o}pez-Cob{\'a}}, {Marino}, {Gil de
  Paz}, {L{\'o}pez-S{\'a}nchez}, {Barrera-Ballesteros}, {Galbany}, {Mast},
  {Abril-Melgarejo}, \& {Roman-Lopes}}]{Sanchez2016a}
{S{\'a}nchez}, S.~F., {P{\'e}rez}, E., {S{\'a}nchez-Bl{\'a}zquez}, P., {et~al.}
  2016{\natexlab{a}}, RMxAA, 52, 171

\bibitem[{{S{\'a}nchez} {et~al.}(2016{\natexlab{b}}){S{\'a}nchez}, {P{\'e}rez},
  {S{\'a}nchez-Bl{\'a}zquez}, {Gonz{\'a}lez}, {Ros{\'a}les-Ortega},
  {Cano-D{\'\i}az}, {L{\'o}pez-Cob{\'a}}, {Marino}, {Gil de Paz}, {Moll{\'a}},
  {L{\'o}pez-S{\'a}nchez}, {Ascasibar}, \&
  {Barrera-Ballesteros}}]{Sanchez2016b}
{S{\'a}nchez}, S.~F., {P{\'e}rez}, E., {S{\'a}nchez-Bl{\'a}zquez}, P., {et~al.}
  2016{\natexlab{b}}, RMxAA, 52, 21

\bibitem[{{S{\'a}nchez Almeida}(2017)}]{SanchezAlmeida2017}
{S{\'a}nchez Almeida}, J. 2017, in Astrophysics and Space Science Library, Vol.
  430, Gas Accretion onto Galaxies, ed. A.~{Fox} \& R.~{Dav{\'e}}, 67

\bibitem[{{S{\'a}nchez Almeida} {et~al.}(2014){S{\'a}nchez Almeida},
  {Elmegreen}, {Mu{\~n}oz-Tu{\~n}{\'o}n}, \& {Elmegreen}}]{SanchezAlmeida2014a}
{S{\'a}nchez Almeida}, J., {Elmegreen}, B.~G., {Mu{\~n}oz-Tu{\~n}{\'o}n}, C.,
  \& {Elmegreen}, D.~M. 2014, A\&ARv, 22, 71

\bibitem[{{S{\'a}nchez-Bl{\'a}zquez} {et~al.}(2011){S{\'a}nchez-Bl{\'a}zquez},
  {Ocvirk}, {Gibson}, {P{\'e}rez}, \& {Peletier}}]{Sanchez-Blazquez2011}
{S{\'a}nchez-Bl{\'a}zquez}, P., {Ocvirk}, P., {Gibson}, B.~K., {P{\'e}rez}, I.,
  \& {Peletier}, R.~F. 2011, MNRAS, 415, 709

\bibitem[{{Sancisi} {et~al.}(2008){Sancisi}, {Fraternali}, {Oosterloo}, \& {van
  der Hulst}}]{Sancisi2008}
{Sancisi}, R., {Fraternali}, F., {Oosterloo}, T., \& {van der Hulst}, T. 2008,
  A\&ARv, 15, 189

\bibitem[{{Sarmiento} {et~al.}(2023){Sarmiento}, {Huertas-Company}, {Knapen},
  {Ibarra-Medel}, {Pillepich}, {S{\'a}nchez}, \& {Boecker}}]{Sarmiento2023}
{Sarmiento}, R., {Huertas-Company}, M., {Knapen}, J.~H., {et~al.} 2023, A\&A,
  673, A23

\bibitem[{{Schaye} {et~al.}(2015){Schaye}, {Crain}, {Bower}, {Furlong},
  {Schaller}, {Theuns}, {Dalla Vecchia}, {Frenk}, {McCarthy}, {Helly},
  {Jenkins}, {Rosas-Guevara}, {White}, {Baes}, {Booth}, {Camps}, {Navarro},
  {Qu}, {Rahmati}, {Sawala}, {Thomas}, \& {Trayford}}]{Schaye2015}
{Schaye}, J., {Crain}, R.~A., {Bower}, R.~G., {et~al.} 2015, MNRAS, 446, 521

\bibitem[{{Schaye} {et~al.}(2010){Schaye}, {Dalla Vecchia}, {Booth}, {Wiersma},
  {Theuns}, {Haas}, {Bertone}, {Duffy}, {McCarthy}, \& {van de
  Voort}}]{Schaye2010}
{Schaye}, J., {Dalla Vecchia}, C., {Booth}, C.~M., {et~al.} 2010, MNRAS, 402,
  1536

\bibitem[{{Schmidt}(1959)}]{Schmidt1959}
{Schmidt}, M. 1959, ApJ, 129, 243

\bibitem[{{Searle} \& {Sargent}(1972)}]{Searle1972}
{Searle}, L. \& {Sargent}, W. L.~W. 1972, ApJ, 173, 25

\bibitem[{{Shen} {et~al.}(2012){Shen}, {Madau}, {Aguirre}, {Guedes}, {Mayer},
  \& {Wadsley}}]{Shen2012}
{Shen}, S., {Madau}, P., {Aguirre}, A., {et~al.} 2012, ApJ, 760, 50

\bibitem[{{Smee} {et~al.}(2013){Smee}, {Gunn}, {Uomoto}, {Roe}, {Schlegel},
  {Rockosi}, {Carr}, {Leger}, {Dawson}, {Olmstead}, {Brinkmann}, {Owen},
  {Barkhouser}, {Honscheid}, {Harding}, {Long}, {Lupton}, {Loomis}, {Anderson},
  {Annis}, {Bernardi}, {Bhardwaj}, {Bizyaev}, {Bolton}, {Brewington}, {Briggs},
  {Burles}, {Burns}, {Castander}, {Connolly}, {Davenport}, {Ebelke}, {Epps},
  {Feldman}, {Friedman}, {Frieman}, {Heckman}, {Hull}, {Knapp}, {Lawrence},
  {Loveday}, {Mannery}, {Malanushenko}, {Malanushenko}, {Merrelli}, {Muna},
  {Newman}, {Nichol}, {Oravetz}, {Pan}, {Pope}, {Ricketts}, {Shelden},
  {Sandford}, {Siegmund}, {Simmons}, {Smith}, {Snedden}, {Schneider},
  {SubbaRao}, {Tremonti}, {Waddell}, \& {York}}]{Smee2013}
{Smee}, S.~A., {Gunn}, J.~E., {Uomoto}, A., {et~al.} 2013, AJ, 146, 32

\bibitem[{{Stark} {et~al.}(2021){Stark}, {Masters}, {Avila-Reese}, {Riffel},
  {Riffel}, {Boardman}, {Zheng}, {Weijmans}, {Dillon}, {Fielder}, {Finnegan},
  {Fofie}, {Goddy}, {Harrington}, {Pace}, {Rujopakarn}, {Samanso}, {Shamsi},
  {Sharma}, {Warrick}, {Witherspoon}, \& {Wolthuis}}]{Stark2021}
{Stark}, D.~V., {Masters}, K.~L., {Avila-Reese}, V., {et~al.} 2021, MNRAS, 503,
  1345

\bibitem[{{Stockinger} {et~al.}(2020){Stockinger}, {Janka}, {Kresse}, {Melson},
  {Ertl}, {Gabler}, {Gessner}, {Wongwathanarat}, {Tolstov}, {Leung}, {Nomoto},
  \& {Heger}}]{Stockinger2020}
{Stockinger}, G., {Janka}, H.~T., {Kresse}, D., {et~al.} 2020, MNRAS, 496, 2039

\bibitem[{{Swinbank} {et~al.}(2012){Swinbank}, {Sobral}, {Smail}, {Geach},
  {Best}, {McCarthy}, {Crain}, \& {Theuns}}]{Swinbank2012}
{Swinbank}, A.~M., {Sobral}, D., {Smail}, I., {et~al.} 2012, MNRAS, 426, 935

\bibitem[{{Tacconi} {et~al.}(2013){Tacconi}, {Neri}, {Genzel}, {Combes},
  {Bolatto}, {Cooper}, {Wuyts}, {Bournaud}, {Burkert}, {Comerford}, {Cox},
  {Davis}, {F{\"o}rster Schreiber}, {Garc{\'\i}a-Burillo}, {Gracia-Carpio},
  {Lutz}, {Naab}, {Newman}, {Omont}, {Saintonge}, {Shapiro Griffin}, {Shapley},
  {Sternberg}, \& {Weiner}}]{Tacconi2013}
{Tacconi}, L.~J., {Neri}, R., {Genzel}, R., {et~al.} 2013, ApJ, 768, 74

\bibitem[{{Tinsley}(1980)}]{Tinsley1980}
{Tinsley}, B.~M. 1980, FCPh, 5, 287

\bibitem[{{Tinsley} \& {Cameron}(1974)}]{Tinsley1974}
{Tinsley}, B.~M. \& {Cameron}, A.~G.~W. 1974, Ap\&SS, 31, 31

\bibitem[{{Torrey} {et~al.}(2019){Torrey}, {Vogelsberger}, {Marinacci},
  {Pakmor}, {Springel}, {Nelson}, {Naiman}, {Pillepich}, {Genel}, {Weinberger},
  \& {Hernquist}}]{Torrey2019}
{Torrey}, P., {Vogelsberger}, M., {Marinacci}, F., {et~al.} 2019, MNRAS, 484,
  5587

\bibitem[{{Tremonti} {et~al.}(2004){Tremonti}, {Heckman}, {Kauffmann},
  {Brinchmann}, {Charlot}, {White}, {Seibert}, {Peng}, {Schlegel}, {Uomoto},
  {Fukugita}, \& {Brinkmann}}]{Tremonti2004}
{Tremonti}, C.~A., {Heckman}, T.~M., {Kauffmann}, G., {et~al.} 2004, ApJ, 613,
  898

\bibitem[{{van de Voort} \& {Schaye}(2012)}]{vandeVoort2012}
{van de Voort}, F. \& {Schaye}, J. 2012, MNRAS, 423, 2991

\bibitem[{{van den Bergh}(1999)}]{VandenBergh1999}
{van den Bergh}, S. 1999, A\&ARv, 9, 273

\bibitem[{{V{\'a}zquez-Mata} {et~al.}(2022){V{\'a}zquez-Mata},
  {Hern{\'a}ndez-Toledo}, {Avila-Reese}, {Herrera-Endoqui},
  {Rodr{\'\i}guez-Puebla}, {Cano-D{\'\i}az}, {Lacerna},
  {Mart{\'\i}nez-V{\'a}zquez}, \& {Lane}}]{Vazquez-Mata2022}
{V{\'a}zquez-Mata}, J.~A., {Hern{\'a}ndez-Toledo}, H.~M., {Avila-Reese}, V.,
  {et~al.} 2022, MNRAS, 512, 2222

\bibitem[{{Veilleux} {et~al.}(2005){Veilleux}, {Cecil}, \&
  {Bland-Hawthorn}}]{Veilleux2005}
{Veilleux}, S., {Cecil}, G., \& {Bland-Hawthorn}, J. 2005, ARA\&A, 43, 769

\bibitem[{{Ventura} {et~al.}(2020){Ventura}, {Dell'Agli}, {Lugaro}, {Romano},
  {Tailo}, \& {Yag{\"u}e}}]{Ventura2020}
{Ventura}, P., {Dell'Agli}, F., {Lugaro}, M., {et~al.} 2020, A\&A, 641, A103

\bibitem[{{Ventura} {et~al.}(2013){Ventura}, {Di Criscienzo}, {Carini}, \&
  {D'Antona}}]{Ventura2013}
{Ventura}, P., {Di Criscienzo}, M., {Carini}, R., \& {D'Antona}, F. 2013,
  MNRAS, 431, 3642

\bibitem[{{Walcher} {et~al.}(2011){Walcher}, {Groves}, {Budav{\'a}ri}, \&
  {Dale}}]{Walcher2011}
{Walcher}, J., {Groves}, B., {Budav{\'a}ri}, T., \& {Dale}, D. 2011, Ap\&SS,
  331, 1

\bibitem[{{Wuyts} {et~al.}(2016){Wuyts}, {Wisnioski}, {Fossati}, {F{\"o}rster
  Schreiber}, {Genzel}, {Davies}, {Mendel}, {Naab}, {R{\"o}ttgers}, {Wilman},
  {Wuyts}, {Bandara}, {Beifiori}, {Belli}, {Bender}, {Brammer}, {Burkert},
  {Chan}, {Galametz}, {Kulkarni}, {Lang}, {Lutz}, {Momcheva}, {Nelson},
  {Rosario}, {Saglia}, {Seitz}, {Tacconi}, {Tadaki}, {{\"U}bler}, \& {van
  Dokkum}}]{Wuyts2016}
{Wuyts}, E., {Wisnioski}, E., {Fossati}, M., {et~al.} 2016, ApJ, 827, 74

\bibitem[{{Yan} {et~al.}(2016){Yan}, {Tremonti}, {Bershady}, {Law}, {Schlegel},
  {Bundy}, {Drory}, {MacDonald}, {Bizyaev}, {Blanc}, {Blanton}, {Cherinka},
  {Eigenbrot}, {Gunn}, {Harding}, {Hogg}, {S{\'a}nchez-Gallego}, {S{\'a}nchez},
  {Wake}, {Weijmans}, {Xiao}, \& {Zhang}}]{Yan2016}
{Yan}, R., {Tremonti}, C., {Bershady}, M.~A., {et~al.} 2016, AJ, 151, 8

\bibitem[{{Yates} {et~al.}(2012){Yates}, {Kauffmann}, \& {Guo}}]{Yates2012}
{Yates}, R.~M., {Kauffmann}, G., \& {Guo}, Q. 2012, MNRAS, 422, 215

\bibitem[{{Young}(1999)}]{Young1999}
{Young}, J.~S. 1999, ApJL, 514, L87

\bibitem[{{Zaragoza-Cardiel} {et~al.}(2019){Zaragoza-Cardiel}, {Fritz},
  {Aretxaga}, {Mayya}, {Rosa-Gonz{\'a}lez}, {Beckman}, {Bruzual}, {Charlot}, \&
  {Lomel{\'\i}-N{\'u}{\~n}ez}}]{Zaragoza-Cardiel2019}
{Zaragoza-Cardiel}, J., {Fritz}, J., {Aretxaga}, I., {et~al.} 2019, MNRAS, 487,
  L61

\bibitem[{{Zhu} {et~al.}(2017){Zhu}, {Barrera-Ballesteros}, {Heckman},
  {Zakamska}, {S{\'a}nchez}, {Yan}, \& {Brinkmann}}]{Zhu2017}
{Zhu}, G.~B., {Barrera-Ballesteros}, J.~K., {Heckman}, T.~M., {et~al.} 2017,
  MNRAS, 468, 4494

\end{thebibliography}


\begin{appendix} 
\section{The effect of the initial gas conditions} \label{sec:appendix}

\begin{figure*}%
\centering
\includegraphics[width=\textwidth]{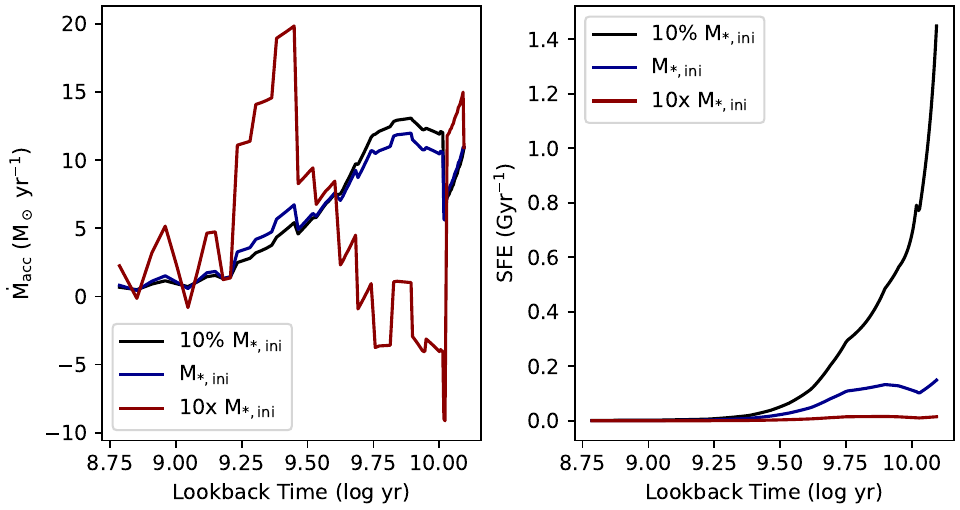}
\caption{Accretion rate (left panel) and SFE (right panel) corresponding to the averaged ChEHs and SFHs of a single stellar mass bin, $10^{11-12}$, using three different values of the initial gas mass. The values are the initial stellar mass, which is used in the main article, and 10\% and 10 times this value.}\label{fig:mprim}
\end{figure*}

The only parameter of the model for which we have no observable constraints is the initial gas mass of the cloud. We reasonably assume from observational data that it is similar to the stellar mass measured at the earliest time, but we need to assess how sensitive the results are to this parameter.
In Fig.~\ref{fig:mprim} we compare the accretion rates and SFE histories obtained using the initial stellar mass with two values separated by an order of magnitude, that is, 10\% and ten times this value.

The accretion rate history appears to change very little when the initial gas mass is lower than our specification, but its shape changes greatly when it is much higher. This is likely due to some sort of "metallicity inertia" such that too large an initial gas mass cannot be sufficiently enriched by the stellar populations, requiring a large negative value, which then requires a similarly large gas mass to dilute it when the abundance is too high. As a result, the values of the accretion rate diverge and fluctuate over time.

In contrast, the course of the SFE is much more strongly influenced by the value of the initial gas mass, especially at the earliest times when the value of the initial gas mass dominates the available gas in the galaxy. The presence of a sharp upturn of the SFE at earlier times can be seen as an indicator that the initial gas mass is underestimated.

\begin{figure*}%
\centering
\includegraphics[width=\textwidth]{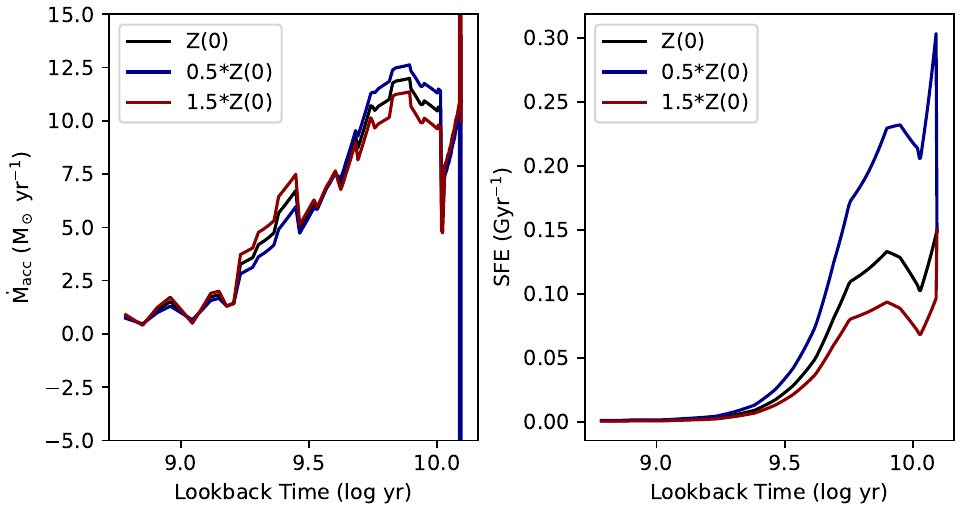}
\caption{Accretion rate (left panel) and SFE (right panel) corresponding to the averaged ChEHs and SFHs of a single stellar mass bin, $10^{11-12}$, using three different values of the iron abundance for the initial gas mass. The values are the measured value for the oldest population as well as half and 1.5 times this value.}\label{fig:zini}
\end{figure*}

The initial gas cloud is also defined by its abundance, the value of which strongly affects the initial accretion rate of the galaxy. We do have a measurement of its value, but unlike the gas mass, the abundance of the oldest population we detected is also the most unreliable data point we have due to the low light content of the high age populations. To assess how critical the precision of this value is in Fig.~\ref{fig:zini}, we show how the shape of the accretion rate and SFE changed when we changed the iron abundance of the initial gas.

The accretion rate is minimally affected, except for the very earliest time when the abundance of ISM is "corrected" by a fast positive or negative accretion to match the measurements. The effect on the SFE is higher because the gas mass difference is maintained over time and affects the SFE values. Since we do not observe the divergence at the earliest times in the SFE, we expect our approach to be consistent.

\end{appendix}
\end{document}